\documentclass[a4paper,twocolumn,english,aps,prl,superscriptaddress]{revtex4-1}
\usepackage[T1]{fontenc}
\usepackage[latin9]{inputenc}
\usepackage{color}
\usepackage{babel}
\usepackage{amsmath}
\usepackage{amssymb}
\usepackage{graphicx}
\DeclareGraphicsExtensions{.pdf}
\usepackage{amsmath,amssymb,bbold,bm,color}
\usepackage{float}
\usepackage{epstopdf}
\usepackage{tabularx}
\usepackage{braket}
\usepackage{booktabs}
\usepackage{array}
\usepackage{blindtext}
\usepackage{natbib}
\usepackage[unicode=true,pdfusetitle,
 bookmarks=true,bookmarksnumbered=false,bookmarksopen=false,
 breaklinks=false,pdfborder={0 0 1},backref=false,colorlinks=false]
 {hyperref}

\makeatletter

\usepackage{babel}
\usepackage{babel}

\begin{document}

\title{Non-Hermitian Quantum Fractals}

\author{Junsong Sun}
\affiliation{School of Physics, Beihang University,
Beijing, 100191, China}

\author{Chang-An Li}
\email{changan.li@uni-wuerzburg.de}
\affiliation{Institute for Theoretical Physics and Astrophysics, University of W$\ddot{u}$rzburg, 97074 W$\ddot{u}$rzburg, Germany}

\author{Qingyang Guo}
\affiliation{School of Physics, Beihang University,
Beijing, 100191, China}

\author{Weixuan Zhang}
\affiliation{Key Laboratory of advanced optoelectronic quantum architecture and measurements of Ministry of Education, Beijing Key Laboratory of Nanophotonics and Ultrafine Optoelectronic Systems, School of Physics, Beijing Institute of Technology, 100081, Beijing, China}

\author{Shiping Feng}
\affiliation{ Department of Physics,  Beijing Normal University, Beijing, 100875, China}

\author{Xiangdong Zhang}
\email{zhangxd@bit.edu.cn}
\affiliation{Key Laboratory of advanced optoelectronic quantum architecture and measurements of Ministry of Education, Beijing Key Laboratory of Nanophotonics and Ultrafine Optoelectronic Systems, School of Physics, Beijing Institute of Technology, 100081, Beijing, China}

\author{Huaiming Guo}
\email{hmguo@buaa.edu.cn}
\affiliation{School of Physics, Beihang University,
Beijing, 100191, China}

\author{Bj$\ddot{\mathrm{o}}$rn Trauzettel}
\email{trauzettel@physik.uni-wuerzburg.de}
\affiliation{Institute for Theoretical Physics and Astrophysics, University of W$\ddot{u}$rzburg, 97074 W$\ddot{u}$rzburg, Germany}

\begin{abstract}
The first quantum fractal discovered in physics is the Hofstadter
butterfly. It stems from large external magnetic fields.
We discover instead a new class of non-Hermitian quantum fractals (NHQFs) emerging in coupled Hatano-Nelson models on a tree lattice in absence of any fields. Based on analytic solutions, we are able to rigorously identify the self-similar
recursive structures in energy spectrum and wave functions. We prove that the complex spectrum of NHQFs bears a resemblance to the Mandelbrot set in fractal theory. The self-similarity of NHQFs is rooted in the interplay between the iterative lattice configuration and non-Hermiticity. Moreover, we show that  NHQFs exist in generalized non-Hermitian systems with iterative lattice structures. Our findings open a new avenue for investigating quantum fractals in non-Hermitian systems. 
\end{abstract}

\date{\today}

\maketitle
\textit{\textcolor{blue}{Introduction.--}}
A fractal describes a geometric pattern with self-similar structures that are ``exactly the same at every scale or nearly the same at different scales''\ \cite{Mandelbrot}.
The prime example of fractals is the Mandelbrot set defined by the simple relation $z_{n+1}=z_{n}^{2}+c$ with self-similar complexity in
the two-dimensional (2D) complex plane.  Fractals have significant impact on
a wide variety of research areas, ranging from mathematics, engineering,
chemistry, to physics\ \cite{Mandelbrot,Fan14nc,Newkome06science,Pietronero,Bunde}.
Hofstadter discovered the connection between fractals and quantum physics in a seminal work\ \cite{Hofstadter76prb}. He showed that 2D Bloch electrons under perpendicular magnetic fields exhibit quantum fractals in the energy-flux plane $(E,\phi)$.
Since then, enormous efforts have explored
the physics of fractals from different aspects including Hofstadter
butterflies\ \cite{Thouless82prl,StredaJPC82,ChangMC96prb,Miyake13prl,Dean13nature},
quantum Hall resistivities\ \cite{PanW03prl}, and Anderson transitions\ \cite{Morgenstern03prl,Evers08rmp,Kosior17prb}.
Renewed interest in the physics of fractals has emerged due to the
rapid progress in topological states of matter\ \cite{Brzezinska18prb,Pai19prb,Fremling20prr,Manna20prr,Iliasov20prb,Biesenthal22science,Ivaki22cp,Manna22prb,Zheng22sb,Stegmaier22prl,LiJ23prr,Stalhammar23prr,Manna23cp,weststrom23arxiv}
and realizations of fractal settings with state of the art techniques\ \cite{Shang15nc,Kempkes19nature,LiuC21prl,Canyellas24nature}. However, few examples of quantum
fractals in the energy spectrum beyond Hofstadter
physics have been predicted so far. 

\begin{figure}[hbpt]
  \includegraphics[width=8.5cm]{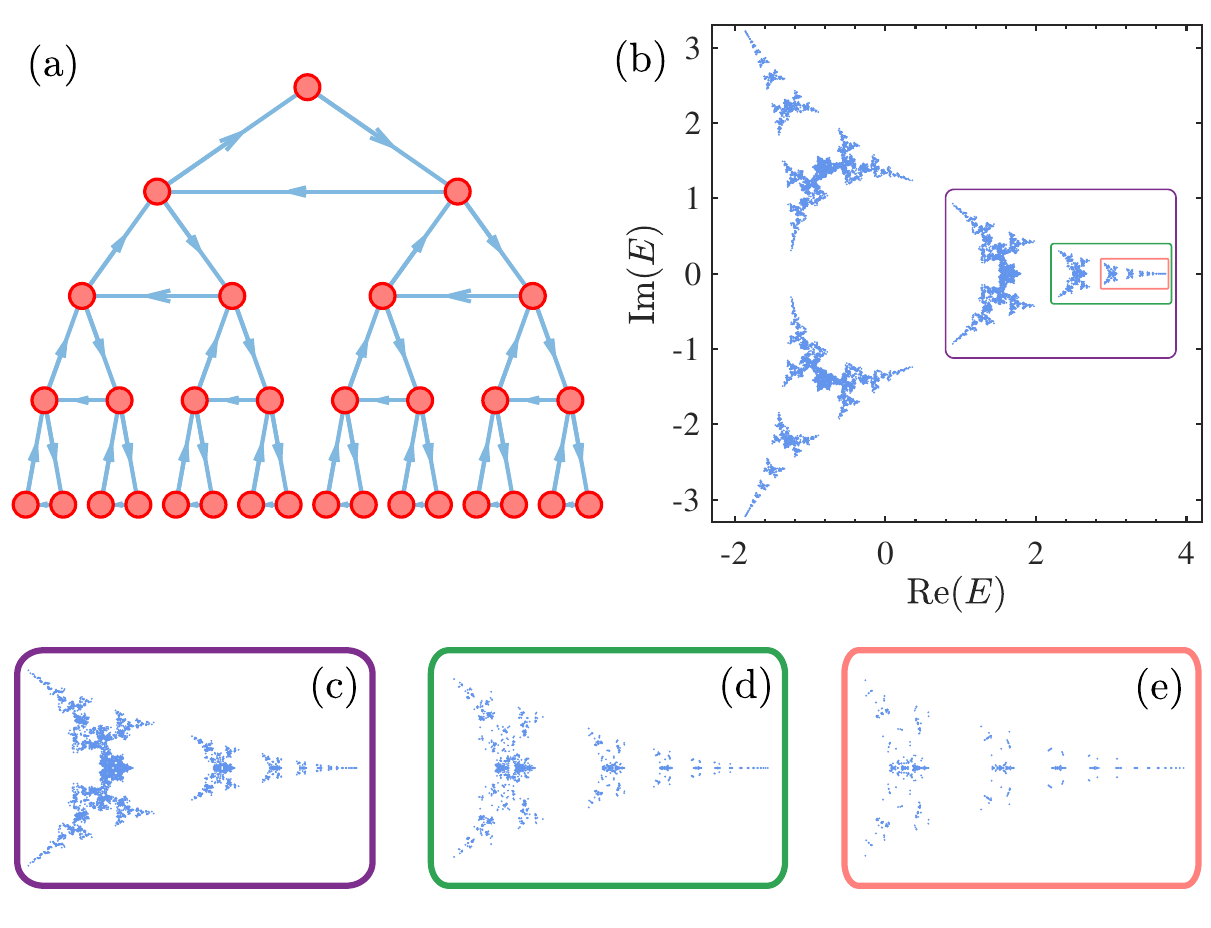}
  \caption{(a) Schematic of the coupled Hatano-Nelson models on a binary tree lattice, where blue arrows represent the non-reciprocal hopping amplitude $t+\gamma$ [the reverse ones with \(t-\gamma\) are omitted for clarity]. (b) The complex energy spectrum corresponding to the model in (a) with $N=16$ generations. (c), (d), and (e) show enlarged plots of the spectrum clusters marked by different colors in panel (b), respectively, revealing a self-similar recursive structure.}\label{fig1}
\end{figure}

Recently, non-Hermitian systems with particular properties have sparked
intense research interests\ \cite{Lee16prl,Yao18prl,Kunst18prl,GonZP18prx,Leykam17prl,LeeCH19prb,ZhangK20prl,ShenH18prl,Okuma20prl,Kawabata19PRL,Yokomizo19prl,Zhou19prb,Borgnia20prl,Budich20prl,Bergholtz21rmp,PhysRevB.108.075122,PRL-LichangAn}.
The complex-valued nature of the
energy spectrum in non-Hermitian systems makes them tantalizing platforms to search for novel quantum fractals in the 2D complex energy plane $(\mathrm{Re}(E),\mathrm{Im}(E))$. Possible realizations of such non-Hermitian quantum
fractals (NHQFs) can be genuine without the need of external fields as for Hofstadter butterflies. 

In this work, we propose a new class of quantum fractals in non-Hermitian systems.  We study exemplarily the properties of coupled Hatano-Nelson models on a tree lattice, as shown in Fig. \ref{fig1}(a). We present
analytic solutions of this non-Hermitian system, which allow us to rigorously
investigate the particular structures of energy spectra and wave functions.
Remarkably, we demonstrate that the electrons exhibit a self-similar recursive energy spectrum in the 2D complex plane {[}see Fig. \ref{fig1}(b){]}.
Its self-similar nature is clearly visible when parts of the spectrum
are magnified step by step {[}see Figs. \ref{fig1}(c, d, e){]}.  We
explain the appearance of NHQFs in terms of the interplay between 
iterative lattice configurations and non-Hermitian effects. Explicitly,
the NHQFs in energy spectra are well described by a variation of the Mandelbrot set in fractal theory. Interestingly, we show that the self-similar recursive character is also present in the structure of wave functions. Finally, we extend the concept of NHQFs to different non-Hermitian systems including the Sierpi$\acute{\mathrm{n}}$ski gasket.

\textit{\textcolor{blue}{Coupled Hatano-Nelson models on a tree lattice.--}}
We start with the coupled Hatano-Nelson models \ \cite{PhysRevLett.77.570,PhysRevB.56.8651} on a tree lattice described by
\begin{equation}
  \mathcal{H}=\sum_{{\langle ij\rangle}}\left[(t-\gamma)c_i^\dagger c_j+(t+\gamma)c_j^\dagger c_i\right],
\end{equation}
where $c_{i}^{\dagger}$ and $c_{i}$ are fermionic creation and annihilation operators at site $i$; $\langle i j\rangle$ denotes nearest-neighbor pairs; $t$ and $\gamma$ represent the Hermitian symmetric and non-Hermitian asymmetric nearest-neighbor hopping amplitudes, respectively. 
The tree lattice is  illustrated in Fig. \ref{fig1}(a), which can effectively be viewed as a two-branch subsystem of the Bethe lattice with a coordination number three. 
Note that the Hatano-Nelson models developed on different branches of the tree are coupled by additional bonds in a particular way such that the non-reciprocal hoppings with  $t+\gamma$ form identical closed loops within each triangle.
The system respects the pseudo-Hermitian symmetry $\eta \mathcal{H}\eta =\mathcal{H}^\dagger$ with $\eta$ a Hermitian invertible operator, which renders the energy spectrum to be real or composed of complex conjugate pairs.

The lattice sites can be classified into different generations as the tree structure grows. Henceforth, the Hamiltonian matrix constructed on different generations exhibits an iterative pattern. For the simplest case with only  the first two generations, the specific Hamiltonian $\mathcal{H}$ in the basis $(c_1,c_2,c_3)^T$ takes the matrix form
\begin{equation}\label{eq2}
  H_{2}=\left(\begin{matrix}
 0&t_+&t_-\\
 t_-&0&t_+\\
 t_+&t_-&0\\
\end{matrix}\right),
\end{equation}
where $t_{\pm}\equiv t\pm\gamma$. For a system with $n>2$ generations, the Hamiltonian $H_n$ in the basis $(c_1,c_2,\cdots,c_{2^n-1})^T$ can be  obtained from $H_{n-1}$ in an iterative way
\begin{equation}\label{eq3}
  H_{n}=\left(\begin{matrix}
 0&t_+{e^T_{n-1}}&t_-e_{n-1}^T\\
 t_-e_{n-1}&H_{n-1}&t_+e_{n-1}{e^T_{n-1}}\\
 t_+e_{n-1}&t_-e_{n-1}e_{n-1}^T&H_{n-1}\\
\end{matrix}\right),
\end{equation}
where $e_n$ is a column vector of dimension $2^n-1$ with elements at the first position being $1$ and zeros otherwise.  $e_n^T$ indicates the transpose of $e_n$. 
The iterative structure of the Hamiltonian arises from the unique tree geometry, as illustrated in Fig. \ref{fig1}(a).

\textit{\textcolor{blue}{Analytic solutions.--}}
The particular iterative structure of the Hamiltonian makes the eigenvalue equation $H_n|\Psi_n\rangle=E|\Psi_n\rangle$ analytically solvable, where $|\Psi_n\rangle=\sum_{i=1}^{2^n-1}\phi_i|i\rangle$ and $|i\rangle=c^\dagger_i|0\rangle$. We sketch the solution below and present details in the Supplemental Materials (SM) \cite{SM}. 
Without loss of generality, we first examine the case $\gamma=t$ corresponding to a unidirectional hopping pattern \cite{Note1}. 
The eigenvalue problem of $H_2$ can be expressed as $H_2|\Psi_2\rangle=E|\Psi_2\rangle$, where the eigenstate is denoted as $|\Psi_2\rangle = \left(\phi_1, \phi_2, \phi_3\right)^T$. 
Solving this eigen-equation, we arrive at  $\epsilon^3\phi_1=\phi_1$ with $\epsilon\equiv E/2t$. Henceforth, the eigenvalues of $H_2$ are roots of the characteristic polynomial equation $P_2(\epsilon) = \epsilon^3 - 1=0.$
The three roots are $\epsilon_2=1$, and $e^{\pm i2\pi/3}$, respectively. 
The eigenstates can be determined correspondingly.

In a similar way, we obtain the eigenvalues of $H_3$ from the characteristic polynomial equation $P_3(\epsilon) = \epsilon P_2^2(\epsilon) - P_1^4(\epsilon) =0$ with $P_1(\epsilon)=\epsilon$. By repeating the above steps, we arrive at a recursive relation for the characteristic polynomial $P_n(\epsilon)$ of the Hamiltonian $H_n$ as
\begin{equation}\label{eq6}
P_n(\epsilon)=\epsilon P_{n-1}^2(\epsilon)-P_{n-2}^4(\epsilon)
\end{equation}
with $n \geq 3$.
The solutions of the equation $P_n (\epsilon)= 0$ yield the energy spectrum of the Hamiltonian $H_n$. 

The corresponding eigenstate of $H_N$ can be written as 
\begin{equation}\label{eq7}
|\Psi_N\rangle=\left(\phi^1_1,\phi^{2}_1,\phi^{2}_2,\cdots,\phi^n_{j},\cdots,\phi^N_1,\cdots,\phi^N_{2^{N-1}}\right)^{T},
\end{equation}
where $\phi^n_{j}$ denotes the components at the $j$-th site of the $n$-th generation and $N$ is the total generation number. In terms of $\phi^N_{j\in \mathrm{odd}}$ located on odd sites in the $N$-th generation,  the even components can be obtained via
\begin{subequations}\label{eq8}
\begin{align}
&\phi^N_{2j}=\epsilon\phi^N_{2(j-1)+1},\\
&\phi^{N-1}_j=\epsilon^2\phi^N_{2(j-1)+1},\\
&\phi^{N-k}_j=\epsilon\phi^{N-k+1}_{2j}-\phi^{N-k+2}_{4(j-1)+3},
\end{align}
\end{subequations}
where $k$ is an integer with the constraint $2<k\leq N$.
The $2^{N-2}$ components \(\phi^N_{2(j-1)+1}\) follow the relation:
\begin{equation}\label{eqphin}
P_{k-1}(\epsilon)\phi^N_{2^{k-1}(j-1)+2^{k-2}-1}=P_{k-2}^2(\epsilon)\phi^N_{2^{k-1}(j-1)+2^{k-1}-1}.
\end{equation}
This yields
\begin{subequations}\label{eq10}
\begin{align}
&\phi^{2}_{j}=\frac{\beta_N^{\pm}}{\alpha_N}\phi_1^1,\  \phi^{3}_j=\frac{\beta^\pm_{N-1}\beta^{\pm}_N}{\alpha_{N-1}\alpha_N }\phi^1_1,\  \cdots, \label{eq10a}\\
&\phi^{N}_j=\frac{\beta_1^{\pm} \cdots\beta^\pm_{N-1}\beta^{\pm}_N }{\alpha_1\cdots\alpha_{N-1}\alpha_N }\phi^1_1,\label{eq10b}
\end{align}
\end{subequations}
where $\alpha_n=P^2_{n-1}(\epsilon)$, $\beta_n^-=4t^2\alpha_{n-1}^2$, and $\beta_n^+=2t\alpha_{n-1}P_{n-1}(\epsilon)$.
The $\pm$ signs in Eqs.~(\ref{eq10a}, \ref{eq10b}) at different sites $j$ are determined by the binary representation of the number $j$, $s_{N}s_{N-1}\cdots s_1$ ($s=0,1$), with $0 (1)$ corresponding to $-(+)$.
Therefore, in conjunction with the normalization condition, the eigenstate $|\Psi_N\rangle$ can be uniquely determined. In such an iterative way, we obtain the exact solution of eigenenergy and eigenstates of the Hamiltonian $H_N$. Note that the solutions for $\gamma\neq t$ can also be obtained following the same procedures \cite{SM}.

\textit{\textcolor{blue}{Non-Hermitian quantum fractals in energy spectra.--}}The energy spectrum of the coupled Hatano-Nelson model on a tree lattice exhibits several interesting features,  as shown in Fig.~\ref{fig1}(b). It has a mirror symmetry with respect to the real axis $\mathrm{Im}(E)=0$ in the complex plane, which stems from the pseudo-Hermitian symmetry of the system. In addition, the energy spectrum shows an emergent three-fold rotation symmetry $C_3$. This comes from the enclosed hopping loops in each triangle of the lattice structure. It is evident from $P_2(\epsilon)=\epsilon^3-1=0$ that if $\epsilon$ is a solution thus the cubic power of $\epsilon$ ensures that $\epsilon e^{\pm 2\pi i/3}$ are also solutions. This property is passed on to the next generation through the iteration relation in Eq.~(\ref{eq6}), resulting in a three-fold rotation symmetry $C_3$ in the whole energy spectrum. It is interesting to see that the $C_3$ symmetry connects the real energy states (stationary states) with complex energy states (corresponding to grow or decay) in an exact way.  The $C_3$ rotation symmetry separates the whole energy spectrum into three different sections. Each section is further divided into an infinite number of spectrum clusters separated by line gaps as the generation number $N$ approaches infinity.

The analytic solutions of $H_n$ obtained above allow us to investigate the fine structure of energy spectra rigorously. Upon careful examination of each section of the spectrum, we find that the energy spectrum exhibits a self-similar recursive pattern, i.e., a NHQF in the complex energy plane. The self-similar nature is clearly visible when parts of the spectrum are magnified step by step as illustrated in Figs. \ref{fig1}(c, d, e). This fractal feature stems from the iterative relation satisfied by the characteristic polynomial of the Hamiltonian, i.e., Eq. (\ref{eq6}).
Indeed, the Hamiltonian $H_n$ in Eq.~(\ref{eq3}) can be interpreted as a modified Mandelbrot matrix\ \cite{afraceigenvector,Chan2017ANK}, as shown in the SM \cite{SM}.  The recursive relation of the characteristic equation described in Eq.~(\ref{eq6})  
represents a variation of the characteristic polynomials of the original Mandelbrot matrix, i.e.,  
\(P_{n+1}(\lambda)=\lambda P_n^2(\lambda)+1 \)\ with a variable $\lambda$ \cite{chaosandfractals,Mandelbrot-like,mandelbrot1980fractal,devaney1999mandelbrot,rochon2000generalized,lei1990similarity}.
As such, NHQFs in the complex spectrum are mathematically described by the Mandelbrot-like set in fractal theory. 

The NHQFs are not inherited from the real-space tree lattices directly since they have different Hausdorff dimensions. Instead, the NHQFs emerge from both non-Hermitian effects and particular tree-lattice configurations. The non-reciprocal hoppings form closed loops on the lattice, thus quantizing the energy levels. The particular tree lattice provides a recursive structure of the discrete energy spectrum. By analogy with the circular motions of particles in Hofstadter physics, the closed non-reciprocal hoppings play the role of magnetic flux. Indeed, similar NHQFs can be obtained if we consider an originally Hermitian model with $\gamma=0$ but applying a magnetic field with an imaginary value \cite{SM}. The imaginary magnetic field renders the system to be non-Hermitian with non-reciprocal hoppings.  However, NHQFs show essential differences from  Hofstadter fractals. The NHQFs result from the interplay between particular non-reciprocal hopping patterns and the recursive lattice structure, distinct from the mechanism responsible for Hofstadter fractals \cite{Note3}.  Besides, NHQFs occur in the complex energy plane $(\mathrm{Re}(E),\mathrm{Im}(E))$, while Hofstadter fractals emerge in the energy-flux plane $(E,\phi)$. Therefore, NHQFs are particular to open quantum systems while Hofstadter fractals exist in closed Hermitian systems. Moreover, NHQFs appear without the need of any external fields. Hence, NHQFs are distinctively different quantum fractals as compared to Hofstadter butterflies.
\begin{figure}[t!]
  \includegraphics[width=8.7cm]{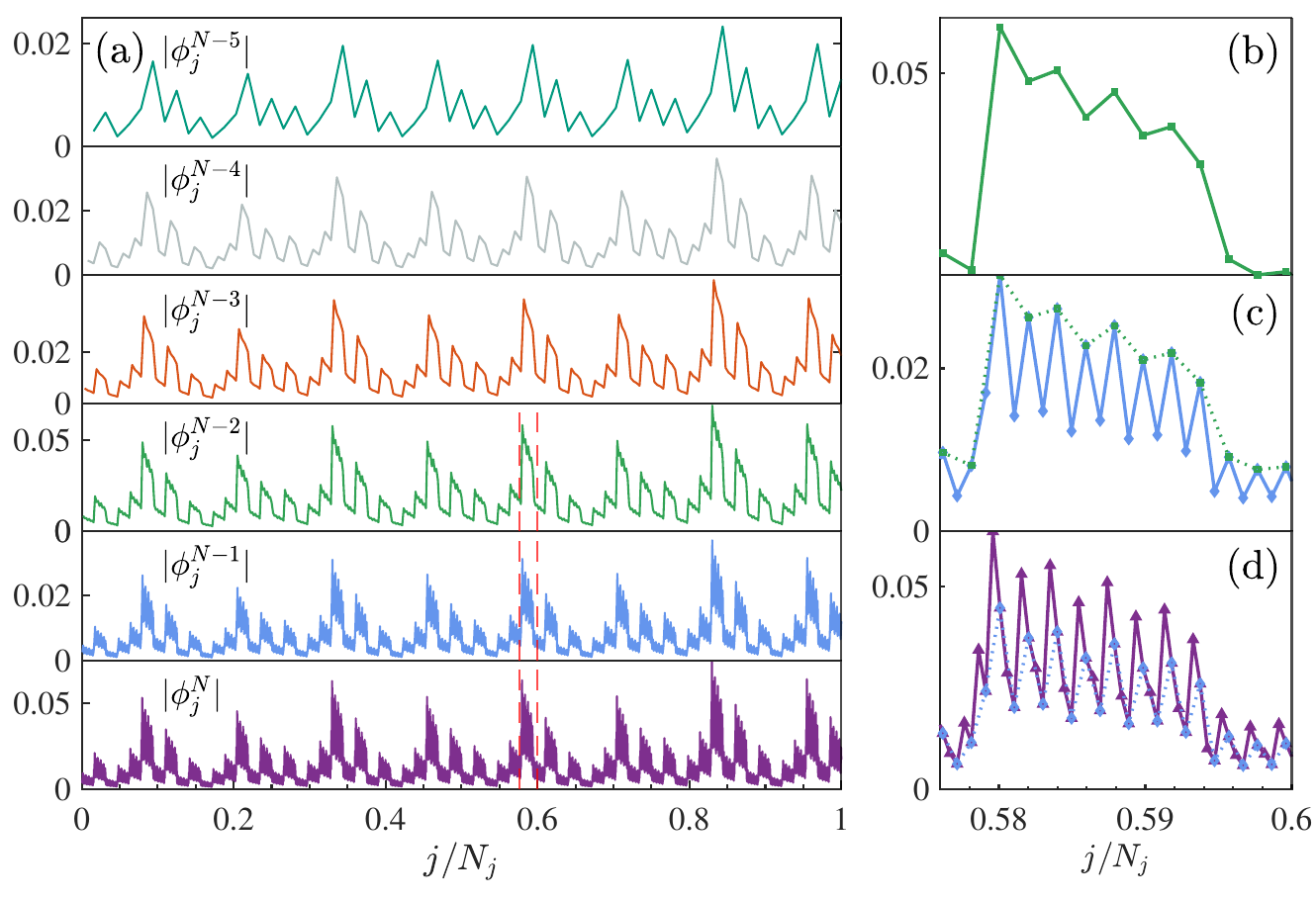}
  \caption{(a) Amplitude of wave functions in each generation as a function of the scaled index $j/N_i$. Here, $N_i=2^{i-1}$ represents the total number of sites in the $i$-th generation. (b), (c), and (d) are enlarged plots of the corresponding small segments in the $(N-2)$-th, $(N-1)$-th, and $N$-th generations, respectively. (c) [(d)] also includes the linearly scaled plot of (b) [(c)], which perfectly matches with each other.}\label{fig2}
\end{figure}

\textit{\textcolor{blue}{Self-similarity in wave functions.--}}
Fractal structures can appear in wave functions as well \cite{PhysRevLett.132.050401,chen2024fractal,wan2024fractal,Rod2012}. Notably,  we demonstrate below that the self-similar recursive patterns of NHQFs appear not only in the energy spectrum, as discussed above, but also in wave functions. Figure \ref{fig2}~(a) plots wave functions at the sites of each generations with respect to a  horizontal length scaled by the factor of its corresponding site number. As the generation number $N$ increases,  the distribution of wave functions takes the same pattern, demonstrating a type of scale invariance. Figure~\ref{fig2}~(b) depicts the amplitude of the wave function in a zoomed window within the corresponding scaled segments of the lowest three generations. Indeed, the curves have precisely the same shape and their amplitudes are equal up to a linear scaling factor.
The self-similarity of wave functions can be recognized by their explicit forms in Eq.~(\ref{eq10}). The value in the \(n\)-th generation is proportional to the ones at odd (even) sites in the \((n+1)\)-th generation with a uniform ratio $\frac{\beta^{-}_{N-n+1}}{\alpha_{N-n+1}}$ ($\frac{\beta^{+}_{N-n+1}}{\alpha_{N-n+1}}$). We find no signatures of non-Hermitian skin effects \cite{Advance-review} by investigating the localization behavior all eigenstates of the system \cite{ReentrantLocalization,LiXiao,LiChang-An}. This is reasonable considering the particular hopping loops in the tree lattices.

\begin{figure}[t!]
  \includegraphics[width=8cm]{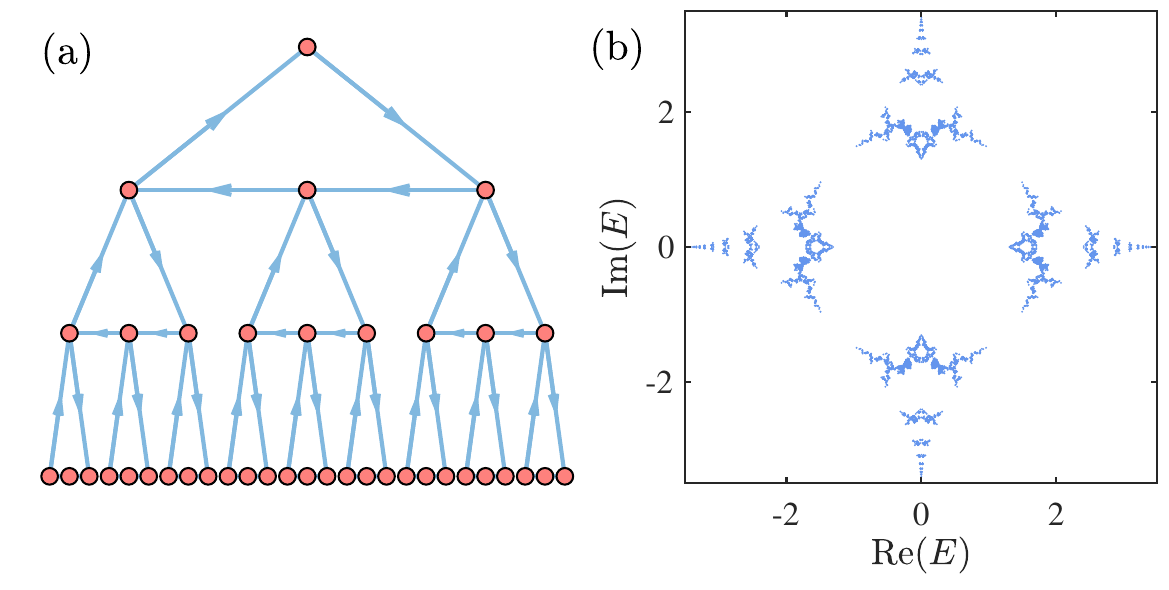}
  \caption{(a) Schematic of a generalized tree lattice to accommodate three branches. (b) Energy spectrum corresponding to extended tree lattices in (a) with $N=10$ generations.}\label{fig3}
\end{figure}

\textit{\textcolor{blue}{Complementary models of non-Hermitian quantum fractals.--}} The guiding principle to search for NHQFs relies on two primary conditions: particular non-reciprocal hopping patterns and iterative lattice structures. As such, NHQFs are expected to be insensitive to specific geometries of the real-space structures as long as the non-Hermitian system exhibits particular iterative structures. Based on this principle, we show below that NHQFs exist in related non-Hermitian systems.  First, we discuss the tree lattice accommodating $q$ branches as another example. We consider a configuration where a parent node only connects two boundary child nodes, forming closed loops in each triangle with non-reciprocal hoppings [see Fig.~\ref{fig3}(a) with $q=3$]. The iterative relation of the characteristic polynomials at $\gamma=t$ is obtained as $P_n(\epsilon)=\epsilon P_{n-1}^q(\epsilon)-P_{n-2}^{q^2}(\epsilon)$ for $(n\geq3)$, where $P_2(\epsilon)=\epsilon^{q+1}-1$.
The $(q+1)$-th power of $\epsilon$ in $P_2(\epsilon)$ implies that the energy spectrum exhibits $(q+1)$-fold rotational symmetry [see Fig.~\ref{fig3}(b) for the $q=3$ case]. Notably, the NHQFs are observed in both the distribution patterns of the energy spectrum and the wave functions.  Furthermore, we can employ squares instead of triangles in the original tree geometries. Due to the common principle, the energy spectrum (as well as wave functions) of this non-Hermitian system presents self-similar recursive  structures as well \cite{SM}.
\begin{figure}[t!]
  \includegraphics[width=8.8cm]{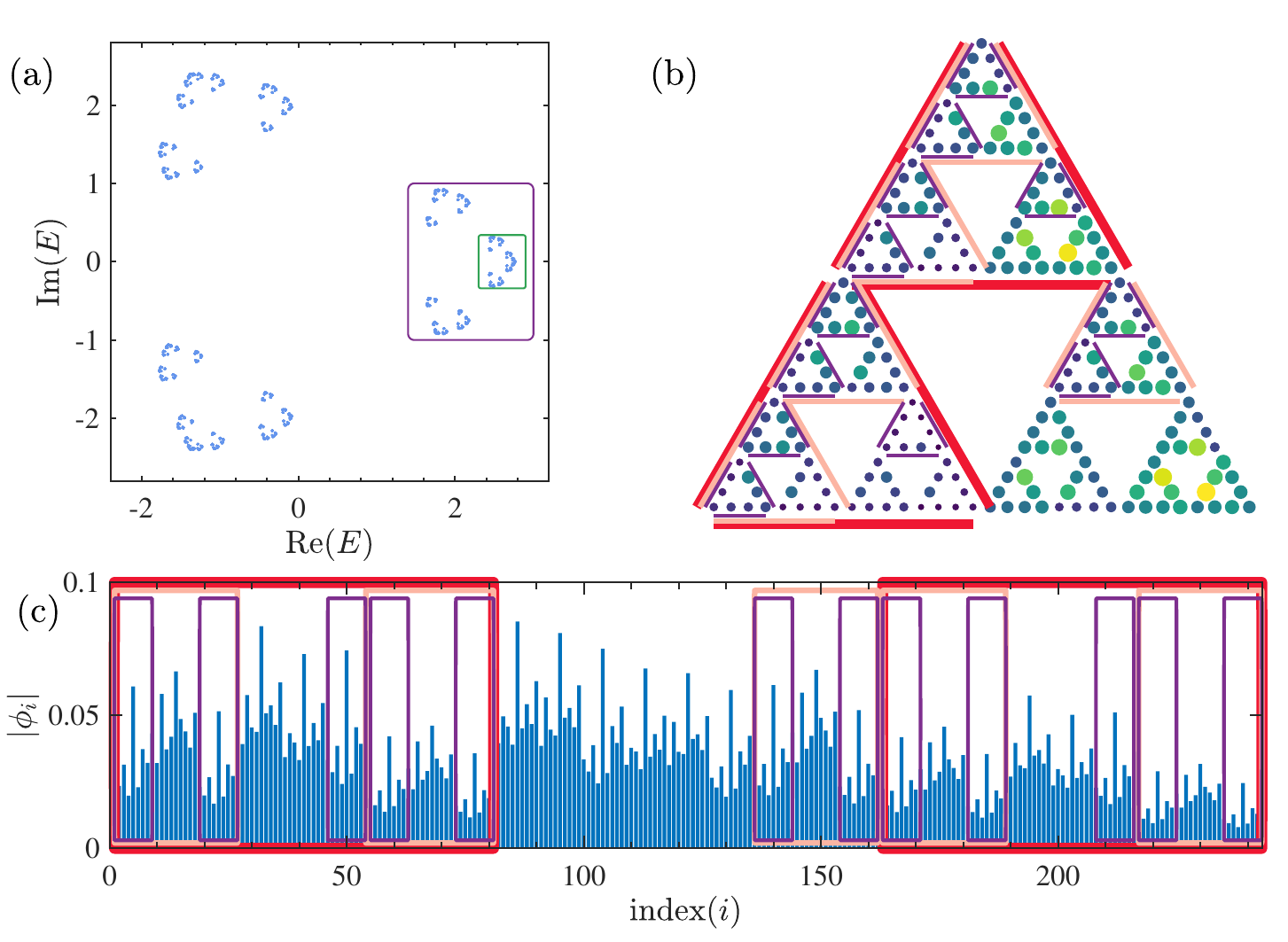}
  \caption{(a) Energy spectrum of coupled Hatano-Nelson models on the Sierpi$\acute{\mathrm{n}}$ski gasket with $N=9$ generations at $t=\gamma$. (b) Schematic demonstration of the wave function of an eigenstate on an $N=6$ lattice with energy $E=-1.2314 + 1.2075{\rm i}$, where the sizes of the markers indicate the magnitudes. We only display the top sub-triangle due to the threefold rotation symmetry. (c) Specific values of the wave function for the eigenstate in (b). Different colored boxes in (c) correspond to the triangular regions of the same colors in (b) marking the similar parts in the wave function.}\label{fig4}
\end{figure}

Finally, we present an example of NHQFs that is not based on the tree geometry. To this end, we consider coupled Hatano-Nelson models on a Sierpi$\acute{\mathrm{n}}$ski gasket [Fig.~\ref{fig4}(b)]. The Sierpi$\acute{\mathrm{n}}$ski gasket also exhibits iterative structures but differs from the tree lattice in important aspects such as the Hausdorff dimension of $d_H \approx 1.58$ and the $C_3$ rotation symmetry of the real-space structure. In this case, the iterative relation to determine the complex energy spectrum becomes \cite{SM}
\begin{equation}
P_n(\epsilon)=P_{n-1}^3(\epsilon)-\prod_{i=0}^{n-3} P_i^6(\epsilon)\left[P_{n-2}^3(\epsilon)-P_{n-1}(\epsilon)\right]^2,
\end{equation}
where $P_0(\epsilon)=\epsilon, P_1(\epsilon)=\epsilon^3-1, P_2(\epsilon)=P_1^3(\epsilon)-1$, $\epsilon=E/2t$, and $n \geq 3$. Hence, the energy spectrum shows a self-similar recursive pattern with global spectrum clusters duplicated at smaller scales, as shown in Fig.~\ref{fig4}(a), which is different from that of the tree-lattice case. Moreover, the self-similarity inherently exists in the wave function of each eigenstate, where the amplitudes within each length scale on the left two triangular segments are linearly correlated. To be specific, the amplitudes of the wave function within the triangular region enclosed by lines of the same color share the same shape, differing only by some proportional coefficients [Figs.~\ref{fig4}(b) and \ref{fig4}(c)]. We can view the Sierpi$\acute{\mathrm{n}}$ski gasket shown in Fig.~\ref{fig4}(b) in terms of three big triangles or equivalently nine sub-triangles. The resemblance of wave functions is observed in sub-triangles \cite{SM}. Note that this resemblance persists in the following generations of the Sierpi$\acute{\mathrm{n}}$ski gasket. 

\textit{\textcolor{blue}{Conclusion.--}}
To summarize, we have discovered NHQFs in coupled Hatano-Nelson models on iterative  lattice structures. The remarkable self-similar recursive features are present in the complex energy spectrum and the wave functions. In terms of the complex energy spectrum, NHQFs are captured by the Mandelbrot-like set in fractal theory. We point to a general principle with a combination of particular non-reciprocal patterns and iterative lattice structures to search for NHQFs in non-Hermitian systems, including the extended tree lattices and Sierpi$\acute{\mathrm{n}}$ski gaskets. We expect that NHQFs can be realized in different physical platforms, such as electric circuits, photonic crystals, and acoustic systems \cite{Lee2018,Imhof2018,Zou2021,ShuoLiu,PhysRevB.103.125411,Nanophoton,PhysRevLett.132.156901,Feng2017,Gao2021,Zhang2021,PhysRevLett.131.066601,PhysRevA.105.023531}. 

\begin{acknowledgments}
The authors thank L.-H. Hu, J. Li, and Z. Wang for
helpful discussions. J.S. and H.G. acknowledges support from the NSFC grant No.~ 12074022. C.A.L. and B.T. were supported by the W$\ddot{\mathrm{u}}$rzburg-Dresden Cluster of Excellence ct.qmat, EXC2147, project-id
390858490, and the DFG (SFB 1170). S.F. is supported by the National Key Research and Development Program of China under Grant Nos. 2023YFA1406500 and 2021YFA1401803, and NSFC under Grant No. 12274036.
\end{acknowledgments}

J. Sun and C.-A. Li contributed equally to this work.


\begin{thebibliography}{46}%
\makeatletter
\providecommand \@ifxundefined [1]{%
 \@ifx{#1\undefined}
}%
\providecommand \@ifnum [1]{%
 \ifnum #1\expandafter \@firstoftwo
 \else \expandafter \@secondoftwo
 \fi
}%
\providecommand \@ifx [1]{%
 \ifx #1\expandafter \@firstoftwo
 \else \expandafter \@secondoftwo
 \fi
}%
\providecommand \natexlab [1]{#1}%
\providecommand \enquote  [1]{``#1''}%
\providecommand \bibnamefont  [1]{#1}%
\providecommand \bibfnamefont [1]{#1}%
\providecommand \citenamefont [1]{#1}%
\providecommand \href@noop [0]{\@secondoftwo}%
\providecommand \href [0]{\begingroup \@sanitize@url \@href}%
\providecommand \@href[1]{\@@startlink{#1}\@@href}%
\providecommand \@@href[1]{\endgroup#1\@@endlink}%
\providecommand \@sanitize@url [0]{\catcode `\\12\catcode `\$12\catcode
  `\&12\catcode `\#12\catcode `\^12\catcode `\_12\catcode `\%12\relax}%
\providecommand \@@startlink[1]{}%
\providecommand \@@endlink[0]{}%
\providecommand \url  [0]{\begingroup\@sanitize@url \@url }%
\providecommand \@url [1]{\endgroup\@href {#1}{\urlprefix }}%
\providecommand \urlprefix  [0]{URL }%
\providecommand \Eprint [0]{\href }%
\providecommand \doibase [0]{http://dx.doi.org/}%
\providecommand \selectlanguage [0]{\@gobble}%
\providecommand \bibinfo  [0]{\@secondoftwo}%
\providecommand \bibfield  [0]{\@secondoftwo}%
\providecommand \translation [1]{[#1]}%
\providecommand \BibitemOpen [0]{}%
\providecommand \bibitemStop [0]{}%
\providecommand \bibitemNoStop [0]{.\EOS\space}%
\providecommand \EOS [0]{\spacefactor3000\relax}%
\providecommand \BibitemShut  [1]{\csname bibitem#1\endcsname}%
\let\auto@bib@innerbib\@empty
\bibitem [{\citenamefont {Mandelbrot}(1982)}]{Mandelbrot}%
  \BibitemOpen
  \bibfield  {author} {\bibinfo {author} {\bibfnamefont {B.~B.}\ \bibnamefont
  {Mandelbrot}},\ }\href@noop {} {\emph {\bibinfo {title} {The Fractal Geometry
  of Nature}}}\ (\bibinfo  {publisher} {W. H. Freeman},\ \bibinfo {address}
  {San Francisco},\ \bibinfo {year} {1982})\BibitemShut {NoStop}%
\bibitem [{\citenamefont {Fan}\ \emph {et~al.}(2014)\citenamefont {Fan},
  \citenamefont {Yeo}, \citenamefont {Su}, \citenamefont {Hattori},
  \citenamefont {Lee}, \citenamefont {Jung}, \citenamefont {Zhang},
  \citenamefont {Liu}, \citenamefont {Cheng}, \citenamefont {Falgout},
  \citenamefont {Bajema}, \citenamefont {Coleman}, \citenamefont {Gregoire},
  \citenamefont {Larsen}, \citenamefont {Huang},\ and\ \citenamefont
  {Rogers}}]{Fan14nc}%
  \BibitemOpen
  \bibfield  {author} {\bibinfo {author} {\bibfnamefont {J.~A.}\ \bibnamefont
  {Fan}}, \bibinfo {author} {\bibfnamefont {W.-H.}\ \bibnamefont {Yeo}},
  \bibinfo {author} {\bibfnamefont {Y.}~\bibnamefont {Su}}, \bibinfo {author}
  {\bibfnamefont {Y.}~\bibnamefont {Hattori}}, \bibinfo {author} {\bibfnamefont
  {W.}~\bibnamefont {Lee}}, \bibinfo {author} {\bibfnamefont {S.-Y.}\
  \bibnamefont {Jung}},  \emph {et~al.},\ }\bibfield  {title} {\enquote
  {\bibinfo {title} {Fractal design concepts for stretchable electronics}},
  }\href {\doibase 10.1038/ncomms4266} {\bibfield  {journal} {\bibinfo
  {journal} {Nat. Commun.}\ }\textbf {\bibinfo {volume} {5}},\
  \bibinfo {pages} {3266} (\bibinfo {year} {2014})}\BibitemShut {NoStop}%
\bibitem [{\citenamefont {Newkome}\ \emph {et~al.}(2006)\citenamefont
  {Newkome}, \citenamefont {Wang}, \citenamefont {Moorefield}, \citenamefont
  {Cho}, \citenamefont {Mohapatra}, \citenamefont {Li}, \citenamefont {Hwang},
  \citenamefont {Lukoyanova}, \citenamefont {Echegoyen}, \citenamefont
  {Palagallo}, \citenamefont {Iancu},\ and\ \citenamefont
  {Hla}}]{Newkome06science}%
  \BibitemOpen
  \bibfield  {author} {\bibinfo {author} {\bibfnamefont {G.~R.}\ \bibnamefont
  {Newkome}}, \bibinfo {author} {\bibfnamefont {P.}~\bibnamefont {Wang}},
  \bibinfo {author} {\bibfnamefont {C.~N.}\ \bibnamefont {Moorefield}},
  \bibinfo {author} {\bibfnamefont {T.~J.}\ \bibnamefont {Cho}}, \bibinfo
  {author} {\bibfnamefont {P.~P.}\ \bibnamefont {Mohapatra}}, \bibinfo {author}
  {\bibfnamefont {S.}~\bibnamefont {Li}},  \emph {et~al.},\ }\bibfield  {title}
  {\enquote {\bibinfo {title} {Nanoassembly of a fractal polymer: A molecular
  "sierpinski hexagonal gasket"}}, }\href {\doibase 10.1126/science.1125894}
  {\bibfield  {journal} {\bibinfo  {journal} {Science}\ }\textbf {\bibinfo
  {volume} {312}},\ \bibinfo {pages} {1782} (\bibinfo {year}
  {2006})}\BibitemShut {NoStop}%
\bibitem [{\citenamefont {Pietronero}\ and\ \citenamefont
  {Tosatti(Eds)}(1986)}]{Pietronero}%
  \BibitemOpen
  \bibfield  {author} {\bibinfo {author} {\bibfnamefont {L.}~\bibnamefont
  {Pietronero}}\ and\ \bibinfo {author} {\bibfnamefont {E.}~\bibnamefont
  {Tosatti(Eds)}},\ }\href@noop {} {\emph {\bibinfo {title} {Fractals in
  Physics}}}\ (\bibinfo  {publisher} {Elsevier},\ \bibinfo {address}
  {Amsterdam},\ \bibinfo {year} {1986})\BibitemShut {NoStop}%
\bibitem [{\citenamefont {Bunde}\ and\ \citenamefont
  {Havlin(Eds)}(1991)}]{Bunde}%
  \BibitemOpen
  \bibfield  {author} {\bibinfo {author} {\bibfnamefont {A.}~\bibnamefont
  {Bunde}}\ and\ \bibinfo {author} {\bibfnamefont {S.}~\bibnamefont
  {Havlin(Eds)}},\ }\href@noop {} {\emph {\bibinfo {title} {Fractals and
  Disordered Systems}}}\ (\bibinfo  {publisher} {Springer},\ \bibinfo {year}
  {1991})\BibitemShut {NoStop}%
\bibitem [{\citenamefont {Hofstadter}(1976)}]{Hofstadter76prb}%
  \BibitemOpen
  \bibfield  {author} {\bibinfo {author} {\bibfnamefont {D.~R.}\ \bibnamefont
  {Hofstadter}},\ }\bibfield  {title} {\enquote {\bibinfo {title} {Energy
  levels and wave functions of bloch electrons in rational and irrational
  magnetic fields}}, }\href {\doibase 10.1103/PhysRevB.14.2239} {\bibfield
  {journal} {\bibinfo  {journal} {Phys. Rev. B}\ }\textbf {\bibinfo {volume}
  {14}},\ \bibinfo {pages} {2239} (\bibinfo {year} {1976})}\BibitemShut
  {NoStop}%
\bibitem [{\citenamefont {Thouless}\ \emph {et~al.}(1982)\citenamefont
  {Thouless}, \citenamefont {Kohmoto}, \citenamefont {Nightingale},\ and\
  \citenamefont {den Nijs}}]{Thouless82prl}%
  \BibitemOpen
  \bibfield  {author} {\bibinfo {author} {\bibfnamefont {D.~J.}\ \bibnamefont
  {Thouless}}, \bibinfo {author} {\bibfnamefont {M.}~\bibnamefont {Kohmoto}},
  \bibinfo {author} {\bibfnamefont {M.~P.}\ \bibnamefont {Nightingale}}, \ and\
  \bibinfo {author} {\bibfnamefont {M.}~\bibnamefont {den Nijs}},\ }\bibfield
  {title} {\enquote {\bibinfo {title} {Quantized hall conductance in a
  two-dimensional periodic potential}}, }\href {\doibase
  10.1103/PhysRevLett.49.405} {\bibfield  {journal} {\bibinfo  {journal} {Phys.
  Rev. Lett.}\ }\textbf {\bibinfo {volume} {49}},\ \bibinfo {pages} {405}
  (\bibinfo {year} {1982})}\BibitemShut {NoStop}%
\bibitem [{\citenamefont {Streda}(1982)}]{StredaJPC82}%
  \BibitemOpen
  \bibfield  {author} {\bibinfo {author} {\bibfnamefont {P.}~\bibnamefont
  {Streda}},\ }\bibfield  {title} {\enquote {\bibinfo {title} {Theory of
  quantised hall conductivity in two dimensions}}, }\href {\doibase
  10.1088/0022-3719/15/22/005} {\bibfield  {journal} {\bibinfo  {journal}
  {J. Phys. C: Solid State Phys.}\ }\textbf {\bibinfo {volume}
  {15}},\ \bibinfo {pages} {L717} (\bibinfo {year} {1982})}\BibitemShut
  {NoStop}%
\bibitem [{\citenamefont {Chang}\ and\ \citenamefont
  {Niu}(1996)}]{ChangMC96prb}%
  \BibitemOpen
  \bibfield  {author} {\bibinfo {author} {\bibfnamefont {M.-C.}\ \bibnamefont
  {Chang}}\ and\ \bibinfo {author} {\bibfnamefont {Q.}~\bibnamefont {Niu}},\
  }\bibfield  {title} {\enquote {\bibinfo {title} {Berry phase, hyperorbits,
  and the hofstadter spectrum: Semiclassical dynamics in magnetic bloch
  bands}}, }\href {\doibase 10.1103/PhysRevB.53.7010} {\bibfield  {journal}
  {\bibinfo  {journal} {Phys. Rev. B}\ }\textbf {\bibinfo {volume} {53}},\
  \bibinfo {pages} {7010} (\bibinfo {year} {1996})}\BibitemShut {NoStop}%
\bibitem [{\citenamefont {Miyake}\ \emph {et~al.}(2013)\citenamefont {Miyake},
  \citenamefont {Siviloglou}, \citenamefont {Kennedy}, \citenamefont {Burton},\
  and\ \citenamefont {Ketterle}}]{Miyake13prl}%
  \BibitemOpen
  \bibfield  {author} {\bibinfo {author} {\bibfnamefont {H.}~\bibnamefont
  {Miyake}}, \bibinfo {author} {\bibfnamefont {G.~A.}\ \bibnamefont
  {Siviloglou}}, \bibinfo {author} {\bibfnamefont {C.~J.}\ \bibnamefont
  {Kennedy}}, \bibinfo {author} {\bibfnamefont {W.~C.}\ \bibnamefont {Burton}},
  \ and\ \bibinfo {author} {\bibfnamefont {W.}~\bibnamefont {Ketterle}},\
  }\bibfield  {title} {\enquote {\bibinfo {title} {Realizing the harper
  hamiltonian with laser-assisted tunneling in optical lattices}}, }\href
  {\doibase 10.1103/PhysRevLett.111.185302} {\bibfield  {journal} {\bibinfo
  {journal} {Phys. Rev. Lett.}\ }\textbf {\bibinfo {volume} {111}},\ \bibinfo
  {pages} {185302} (\bibinfo {year} {2013})}\BibitemShut {NoStop}%
\bibitem [{\citenamefont {Dean}\ \emph {et~al.}(2013)\citenamefont {Dean},
  \citenamefont {Wang}, \citenamefont {Maher}, \citenamefont {Forsythe},
  \citenamefont {Ghahari}, \citenamefont {Gao}, \citenamefont {Katoch},
  \citenamefont {Ishigami}, \citenamefont {Moon}, \citenamefont {Koshino},
  \citenamefont {Taniguchi}, \citenamefont {Watanabe}, \citenamefont {Shepard},
  \citenamefont {Hone},\ and\ \citenamefont {Kim}}]{Dean13nature}%
  \BibitemOpen
  \bibfield  {author} {\bibinfo {author} {\bibfnamefont {C.~R.}\ \bibnamefont
  {Dean}}, \bibinfo {author} {\bibfnamefont {L.}~\bibnamefont {Wang}}, \bibinfo
  {author} {\bibfnamefont {P.}~\bibnamefont {Maher}}, \bibinfo {author}
  {\bibfnamefont {C.}~\bibnamefont {Forsythe}}, \bibinfo {author}
  {\bibfnamefont {F.}~\bibnamefont {Ghahari}}, \bibinfo {author} {\bibfnamefont
  {Y.}~\bibnamefont {Gao}},  \emph {et~al.},\ }\bibfield  {title} {\enquote
  {\bibinfo {title} {Hofstadter's butterfly and the fractal quantum hall effect
  in moir{\'e} superlattices}}, }\href {\doibase 10.1038/nature12186}
  {\bibfield  {journal} {\bibinfo  {journal} {Nature}\ }\textbf {\bibinfo
  {volume} {497}},\ \bibinfo {pages} {598} (\bibinfo {year}
  {2013})}\BibitemShut {NoStop}%
\bibitem [{\citenamefont {Pan}\ \emph {et~al.}(2003)\citenamefont {Pan},
  \citenamefont {Stormer}, \citenamefont {Tsui}, \citenamefont {Pfeiffer},
  \citenamefont {Baldwin},\ and\ \citenamefont {West}}]{PanW03prl}%
  \BibitemOpen
  \bibfield  {author} {\bibinfo {author} {\bibfnamefont {W.}~\bibnamefont
  {Pan}}, \bibinfo {author} {\bibfnamefont {H.~L.}\ \bibnamefont {Stormer}},
  \bibinfo {author} {\bibfnamefont {D.~C.}\ \bibnamefont {Tsui}}, \bibinfo
  {author} {\bibfnamefont {L.~N.}\ \bibnamefont {Pfeiffer}}, \bibinfo {author}
  {\bibfnamefont {K.~W.}\ \bibnamefont {Baldwin}}, \ and\ \bibinfo {author}
  {\bibfnamefont {K.~W.}\ \bibnamefont {West}},\ }\bibfield  {title} {\enquote
  {\bibinfo {title} {Fractional quantum hall effect of composite fermions}},
  }\href {\doibase 10.1103/PhysRevLett.90.016801} {\bibfield  {journal}
  {\bibinfo  {journal} {Phys. Rev. Lett.}\ }\textbf {\bibinfo {volume} {90}},\
  \bibinfo {pages} {016801} (\bibinfo {year} {2003})}\BibitemShut {NoStop}%
\bibitem [{\citenamefont {Morgenstern}\ \emph {et~al.}(2003)\citenamefont
  {Morgenstern}, \citenamefont {Klijn}, \citenamefont {Meyer},\ and\
  \citenamefont {Wiesendanger}}]{Morgenstern03prl}%
  \BibitemOpen
  \bibfield  {author} {\bibinfo {author} {\bibfnamefont {M.}~\bibnamefont
  {Morgenstern}}, \bibinfo {author} {\bibfnamefont {J.}~\bibnamefont {Klijn}},
  \bibinfo {author} {\bibfnamefont {C.}~\bibnamefont {Meyer}}, \ and\ \bibinfo
  {author} {\bibfnamefont {R.}~\bibnamefont {Wiesendanger}},\ }\bibfield
  {title} {\enquote {\bibinfo {title} {Real-space observation of drift states
  in a two-dimensional electron system at high magnetic fields}}, }\href
  {\doibase 10.1103/PhysRevLett.90.056804} {\bibfield  {journal} {\bibinfo
  {journal} {Phys. Rev. Lett.}\ }\textbf {\bibinfo {volume} {90}},\ \bibinfo
  {pages} {056804} (\bibinfo {year} {2003})}\BibitemShut {NoStop}%
\bibitem [{\citenamefont {Evers}\ and\ \citenamefont
  {Mirlin}(2008)}]{Evers08rmp}%
  \BibitemOpen
  \bibfield  {author} {\bibinfo {author} {\bibfnamefont {F.}~\bibnamefont
  {Evers}}\ and\ \bibinfo {author} {\bibfnamefont {A.~D.}\ \bibnamefont
  {Mirlin}},\ }\bibfield  {title} {\enquote {\bibinfo {title} {Anderson
  transitions}}, }\href {\doibase 10.1103/RevModPhys.80.1355} {\bibfield
  {journal} {\bibinfo  {journal} {Rev. Mod. Phys.}\ }\textbf {\bibinfo {volume}
  {80}},\ \bibinfo {pages} {1355} (\bibinfo {year} {2008})}\BibitemShut
  {NoStop}%
\bibitem [{\citenamefont {Kosior}\ and\ \citenamefont
  {Sacha}(2017)}]{Kosior17prb}%
  \BibitemOpen
  \bibfield  {author} {\bibinfo {author} {\bibfnamefont {A.}~\bibnamefont
  {Kosior}}\ and\ \bibinfo {author} {\bibfnamefont {K.}~\bibnamefont {Sacha}},\
  }\bibfield  {title} {\enquote {\bibinfo {title} {Localization in random
  fractal lattices}}, }\href {\doibase 10.1103/PhysRevB.95.104206} {\bibfield
  {journal} {\bibinfo  {journal} {Phys. Rev. B}\ }\textbf {\bibinfo {volume}
  {95}},\ \bibinfo {pages} {104206} (\bibinfo {year} {2017})}\BibitemShut
  {NoStop}%
\bibitem [{\citenamefont {Brzezi\ifmmode~\acute{n}\else \'{n}\fi{}ska}\ \emph
  {et~al.}(2018)\citenamefont {Brzezi\ifmmode~\acute{n}\else \'{n}\fi{}ska},
  \citenamefont {Cook},\ and\ \citenamefont {Neupert}}]{Brzezinska18prb}%
  \BibitemOpen
  \bibfield  {author} {\bibinfo {author} {\bibfnamefont {M.}~\bibnamefont
  {Brzezi\ifmmode~\acute{n}\else \'{n}\fi{}ska}}, \bibinfo {author}
  {\bibfnamefont {A.~M.}\ \bibnamefont {Cook}}, \ and\ \bibinfo {author}
  {\bibfnamefont {T.}~\bibnamefont {Neupert}},\ }\bibfield  {title} {\enquote
  {\bibinfo {title} {Topology in the sierpi\ifmmode \acute{n}\else
  \'{n}\fi{}ski-hofstadter problem}}, }\href {\doibase
  10.1103/PhysRevB.98.205116} {\bibfield  {journal} {\bibinfo  {journal} {Phys.
  Rev. B}\ }\textbf {\bibinfo {volume} {98}},\ \bibinfo {pages} {205116}
  (\bibinfo {year} {2018})}\BibitemShut {NoStop}%
\bibitem [{\citenamefont {Pai}\ and\ \citenamefont {Prem}(2019)}]{Pai19prb}%
  \BibitemOpen
  \bibfield  {author} {\bibinfo {author} {\bibfnamefont {S.}~\bibnamefont
  {Pai}}\ and\ \bibinfo {author} {\bibfnamefont {A.}~\bibnamefont {Prem}},\
  }\bibfield  {title} {\enquote {\bibinfo {title} {Topological states on
  fractal lattices}}, }\href {\doibase 10.1103/PhysRevB.100.155135} {\bibfield
  {journal} {\bibinfo  {journal} {Phys. Rev. B}\ }\textbf {\bibinfo {volume}
  {100}},\ \bibinfo {pages} {155135} (\bibinfo {year} {2019})}\BibitemShut
  {NoStop}%
\bibitem [{\citenamefont {Fremling}\ \emph {et~al.}(2020)\citenamefont
  {Fremling}, \citenamefont {van Hooft}, \citenamefont {Smith},\ and\
  \citenamefont {Fritz}}]{Fremling20prr}%
  \BibitemOpen
  \bibfield  {author} {\bibinfo {author} {\bibfnamefont {M.}~\bibnamefont
  {Fremling}}, \bibinfo {author} {\bibfnamefont {M.}~\bibnamefont {van Hooft}},
  \bibinfo {author} {\bibfnamefont {C.~M.}\ \bibnamefont {Smith}}, \ and\
  \bibinfo {author} {\bibfnamefont {L.}~\bibnamefont {Fritz}},\ }\bibfield
  {title} {\enquote {\bibinfo {title} {Existence of robust edge currents in
  sierpi\ifmmode \acute{n}\else \'{n}\fi{}ski fractals}}, }\href {\doibase
  10.1103/PhysRevResearch.2.013044} {\bibfield  {journal} {\bibinfo  {journal}
  {Phys. Rev. Res.}\ }\textbf {\bibinfo {volume} {2}},\ \bibinfo {pages}
  {013044} (\bibinfo {year} {2020})}\BibitemShut {NoStop}%
\bibitem [{\citenamefont {Manna}\ \emph {et~al.}(2020)\citenamefont {Manna},
  \citenamefont {Pal}, \citenamefont {Wang},\ and\ \citenamefont
  {Nielsen}}]{Manna20prr}%
  \BibitemOpen
  \bibfield  {author} {\bibinfo {author} {\bibfnamefont {S.}~\bibnamefont
  {Manna}}, \bibinfo {author} {\bibfnamefont {B.}~\bibnamefont {Pal}}, \bibinfo
  {author} {\bibfnamefont {W.}~\bibnamefont {Wang}}, \ and\ \bibinfo {author}
  {\bibfnamefont {A.~E.~B.}\ \bibnamefont {Nielsen}},\ }\bibfield  {title}
  {\enquote {\bibinfo {title} {Anyons and fractional quantum hall effect in
  fractal dimensions}}, }\href {\doibase 10.1103/PhysRevResearch.2.023401}
  {\bibfield  {journal} {\bibinfo  {journal} {Phys. Rev. Res.}\ }\textbf
  {\bibinfo {volume} {2}},\ \bibinfo {pages} {023401} (\bibinfo {year}
  {2020})}\BibitemShut {NoStop}%
\bibitem [{\citenamefont {Iliasov}\ \emph {et~al.}(2020)\citenamefont
  {Iliasov}, \citenamefont {Katsnelson},\ and\ \citenamefont
  {Yuan}}]{Iliasov20prb}%
  \BibitemOpen
  \bibfield  {author} {\bibinfo {author} {\bibfnamefont {A.~A.}\ \bibnamefont
  {Iliasov}}, \bibinfo {author} {\bibfnamefont {M.~I.}\ \bibnamefont
  {Katsnelson}}, \ and\ \bibinfo {author} {\bibfnamefont {S.}~\bibnamefont
  {Yuan}},\ }\bibfield  {title} {\enquote {\bibinfo {title} {Hall conductivity
  of a sierpi\ifmmode \acute{n}\else \'{n}\fi{}ski carpet}}, }\href {\doibase
  10.1103/PhysRevB.101.045413} {\bibfield  {journal} {\bibinfo  {journal}
  {Phys. Rev. B}\ }\textbf {\bibinfo {volume} {101}},\ \bibinfo {pages}
  {045413} (\bibinfo {year} {2020})}\BibitemShut {NoStop}%
\bibitem [{\citenamefont {Biesenthal}\ \emph {et~al.}(2022)\citenamefont
  {Biesenthal}, \citenamefont {Maczewsky}, \citenamefont {Yang}, \citenamefont
  {Kremer}, \citenamefont {Segev}, \citenamefont {Szameit},\ and\ \citenamefont
  {Heinrich}}]{Biesenthal22science}%
  \BibitemOpen
  \bibfield  {author} {\bibinfo {author} {\bibfnamefont {T.}~\bibnamefont
  {Biesenthal}}, \bibinfo {author} {\bibfnamefont {L.~J.}\ \bibnamefont
  {Maczewsky}}, \bibinfo {author} {\bibfnamefont {Z.}~\bibnamefont {Yang}},
  \bibinfo {author} {\bibfnamefont {M.}~\bibnamefont {Kremer}}, \bibinfo
  {author} {\bibfnamefont {M.}~\bibnamefont {Segev}}, \bibinfo {author}
  {\bibfnamefont {A.}~\bibnamefont {Szameit}}, \ and\ \bibinfo {author}
  {\bibfnamefont {M.}~\bibnamefont {Heinrich}},\ }\bibfield  {title} {\enquote
  {\bibinfo {title} {Fractal photonic topological insulators}}, }\href
  {\doibase 10.1126/science.abm2842} {\bibfield  {journal} {\bibinfo  {journal}
  {Science}\ }\textbf {\bibinfo {volume} {376}},\ \bibinfo {pages} {1114}
  (\bibinfo {year} {2022})}\BibitemShut
  {NoStop}%
\bibitem [{\citenamefont {Ivaki}\ \emph {et~al.}(2022)\citenamefont {Ivaki},
  \citenamefont {Sahlberg}, \citenamefont {P{\"o}yh{\"o}nen},\ and\
  \citenamefont {Ojanen}}]{Ivaki22cp}%
  \BibitemOpen
  \bibfield  {author} {\bibinfo {author} {\bibfnamefont {M.~N.}\ \bibnamefont
  {Ivaki}}, \bibinfo {author} {\bibfnamefont {I.}~\bibnamefont {Sahlberg}},
  \bibinfo {author} {\bibfnamefont {K.}~\bibnamefont {P{\"o}yh{\"o}nen}}, \
  and\ \bibinfo {author} {\bibfnamefont {T.}~\bibnamefont {Ojanen}},\
  }\bibfield  {title} {\enquote {\bibinfo {title} {Topological random
  fractals}}, }\href {\doibase 10.1038/s42005-022-01101-z} {\bibfield
  {journal} {\bibinfo  {journal} {Commun. Phys.}\ }\textbf {\bibinfo
  {volume} {5}},\ \bibinfo {pages} {327} (\bibinfo {year} {2022})}\BibitemShut
  {NoStop}%
\bibitem [{\citenamefont {Manna}\ \emph {et~al.}(2022)\citenamefont {Manna},
  \citenamefont {Nandy},\ and\ \citenamefont {Roy}}]{Manna22prb}%
  \BibitemOpen
  \bibfield  {author} {\bibinfo {author} {\bibfnamefont {S.}~\bibnamefont
  {Manna}}, \bibinfo {author} {\bibfnamefont {S.}~\bibnamefont {Nandy}}, \ and\
  \bibinfo {author} {\bibfnamefont {B.}~\bibnamefont {Roy}},\ }\bibfield
  {title} {\enquote {\bibinfo {title} {Higher-order topological phases on
  fractal lattices}}, }\href {\doibase 10.1103/PhysRevB.105.L201301} {\bibfield
   {journal} {\bibinfo  {journal} {Phys. Rev. B}\ }\textbf {\bibinfo {volume}
  {105}},\ \bibinfo {pages} {L201301} (\bibinfo {year} {2022})}\BibitemShut
  {NoStop}%
\bibitem [{\citenamefont {Zheng}\ \emph {et~al.}(2022)\citenamefont {Zheng},
  \citenamefont {Man}, \citenamefont {Kong}, \citenamefont {Lin}, \citenamefont
  {Duan}, \citenamefont {Chen}, \citenamefont {Yu}, \citenamefont {Jiang},\
  and\ \citenamefont {Xia}}]{Zheng22sb}%
  \BibitemOpen
  \bibfield  {author} {\bibinfo {author} {\bibfnamefont {S.}~\bibnamefont
  {Zheng}}, \bibinfo {author} {\bibfnamefont {X.}~\bibnamefont {Man}}, \bibinfo
  {author} {\bibfnamefont {Z.-L.}\ \bibnamefont {Kong}}, \bibinfo {author}
  {\bibfnamefont {Z.-K.}\ \bibnamefont {Lin}}, \bibinfo {author} {\bibfnamefont
  {G.}~\bibnamefont {Duan}}, \bibinfo {author} {\bibfnamefont {N.}~\bibnamefont
  {Chen}}, \bibinfo {author} {\bibfnamefont {D.}~\bibnamefont {Yu}}, \bibinfo
  {author} {\bibfnamefont {J.-H.}\ \bibnamefont {Jiang}}, \ and\ \bibinfo
  {author} {\bibfnamefont {B.}~\bibnamefont {Xia}},\ }\bibfield  {title}
  {\enquote {\bibinfo {title} {Observation of fractal higher-order topological
  states in acoustic metamaterials}}, }\href {\doibase
  https://doi.org/10.1016/j.scib.2022.09.020} {\bibfield  {journal} {\bibinfo
  {journal} {Sci. Bull.}\ }\textbf {\bibinfo {volume} {67}},\ \bibinfo
  {pages} {2069} (\bibinfo {year} {2022})}\BibitemShut {NoStop}%
\bibitem [{\citenamefont {Stegmaier}\ \emph {et~al.}(2022)\citenamefont
  {Stegmaier}, \citenamefont {Upreti}, \citenamefont {Thomale},\ and\
  \citenamefont {Boettcher}}]{Stegmaier22prl}%
  \BibitemOpen
  \bibfield  {author} {\bibinfo {author} {\bibfnamefont {A.}~\bibnamefont
  {Stegmaier}}, \bibinfo {author} {\bibfnamefont {L.~K.}\ \bibnamefont
  {Upreti}}, \bibinfo {author} {\bibfnamefont {R.}~\bibnamefont {Thomale}}, \
  and\ \bibinfo {author} {\bibfnamefont {I.}~\bibnamefont {Boettcher}},\
  }\bibfield  {title} {\enquote {\bibinfo {title} {Universality of hofstadter
  butterflies on hyperbolic lattices}}, }\href {\doibase
  10.1103/PhysRevLett.128.166402} {\bibfield  {journal} {\bibinfo  {journal}
  {Phys. Rev. Lett.}\ }\textbf {\bibinfo {volume} {128}},\ \bibinfo {pages}
  {166402} (\bibinfo {year} {2022})}\BibitemShut {NoStop}%
\bibitem [{\citenamefont {Li}\ \emph {et~al.}(2023)\citenamefont {Li},
  \citenamefont {Sun}, \citenamefont {Mo}, \citenamefont {Ruan},\ and\
  \citenamefont {Yang}}]{LiJ23prr}%
  \BibitemOpen
  \bibfield  {author} {\bibinfo {author} {\bibfnamefont {J.}~\bibnamefont
  {Li}}, \bibinfo {author} {\bibfnamefont {Y.}~\bibnamefont {Sun}}, \bibinfo
  {author} {\bibfnamefont {Q.}~\bibnamefont {Mo}}, \bibinfo {author}
  {\bibfnamefont {Z.}~\bibnamefont {Ruan}}, \ and\ \bibinfo {author}
  {\bibfnamefont {Z.}~\bibnamefont {Yang}},\ }\bibfield  {title} {\enquote
  {\bibinfo {title} {Fractality-induced topological phase squeezing and devil's
  staircase}}, }\href {\doibase 10.1103/PhysRevResearch.5.023189} {\bibfield
  {journal} {\bibinfo  {journal} {Phys. Rev. Res.}\ }\textbf {\bibinfo {volume}
  {5}},\ \bibinfo {pages} {023189} (\bibinfo {year} {2023})}\BibitemShut
  {NoStop}%
\bibitem [{\citenamefont {St\aa{}lhammar}\ and\ \citenamefont
  {Morais~Smith}(2023)}]{Stalhammar23prr}%
  \BibitemOpen
  \bibfield  {author} {\bibinfo {author} {\bibfnamefont {M.}~\bibnamefont
  {St\aa{}lhammar}}\ and\ \bibinfo {author} {\bibfnamefont {C.}~\bibnamefont
  {Morais~Smith}},\ }\bibfield  {title} {\enquote {\bibinfo {title} {Fractal
  nodal band structures}}, }\href {\doibase 10.1103/PhysRevResearch.5.043043}
  {\bibfield  {journal} {\bibinfo  {journal} {Phys. Rev. Res.}\ }\textbf
  {\bibinfo {volume} {5}},\ \bibinfo {pages} {043043} (\bibinfo {year}
  {2023})}\BibitemShut {NoStop}%
\bibitem [{\citenamefont {Manna}\ and\ \citenamefont {Roy}(2023)}]{Manna23cp}%
  \BibitemOpen
  \bibfield  {author} {\bibinfo {author} {\bibfnamefont {S.}~\bibnamefont
  {Manna}}\ and\ \bibinfo {author} {\bibfnamefont {B.}~\bibnamefont {Roy}},\
  }\bibfield  {title} {\enquote {\bibinfo {title} {Inner skin effects on
  non-hermitian topological fractals}}, }\href {\doibase
  10.1038/s42005-023-01130-2} {\bibfield  {journal} {\bibinfo  {journal}
  {Commun. Phys.}\ }\textbf {\bibinfo {volume} {6}},\ \bibinfo {pages}
  {10} (\bibinfo {year} {2023})}\BibitemShut {NoStop}%
\bibitem [{\citenamefont {Westström}\ \emph {et~al.}(2023)\citenamefont
  {Westström}, \citenamefont {Duan}, \citenamefont {Yao}, \citenamefont
  {Wang}, \citenamefont {Liu},\ and\ \citenamefont {Li}}]{weststrom23arxiv}%
  \BibitemOpen
  \bibfield  {author} {\bibinfo {author} {\bibfnamefont {A.}~\bibnamefont
  {Weststrom}}, \bibinfo {author} {\bibfnamefont {W.}~\bibnamefont {Duan}},
  \bibinfo {author} {\bibfnamefont {K.}~\bibnamefont {Yao}}, \bibinfo {author}
  {\bibfnamefont {X.}~\bibnamefont {Wang}}, \bibinfo {author} {\bibfnamefont
  {J.}~\bibnamefont {Liu}}, \ and\ \bibinfo {author} {\bibfnamefont
  {J.}~\bibnamefont {Li}},\ }\href@noop {} {\enquote {\bibinfo {title}
  {Topological phases on quantum trees}}, } (\bibinfo {year} {2023}),\ \Eprint
  {http://arxiv.org/abs/2302.03166} {arXiv:2302.03166 [cond-mat.mes-hall]}
  \BibitemShut {NoStop}%
\bibitem [{\citenamefont {Shang}\ \emph {et~al.}(2015)\citenamefont {Shang},
  \citenamefont {Wang}, \citenamefont {Chen}, \citenamefont {Dai},
  \citenamefont {Zhou}, \citenamefont {Kuttner}, \citenamefont {Hilt},
  \citenamefont {Shao}, \citenamefont {Gottfried},\ and\ \citenamefont
  {Wu}}]{Shang15nc}%
  \BibitemOpen
  \bibfield  {author} {\bibinfo {author} {\bibfnamefont {J.}~\bibnamefont
  {Shang}}, \bibinfo {author} {\bibfnamefont {Y.}~\bibnamefont {Wang}},
  \bibinfo {author} {\bibfnamefont {M.}~\bibnamefont {Chen}}, \bibinfo {author}
  {\bibfnamefont {J.}~\bibnamefont {Dai}}, \bibinfo {author} {\bibfnamefont
  {X.}~\bibnamefont {Zhou}}, \bibinfo {author} {\bibfnamefont {J.}~\bibnamefont
  {Kuttner}}, \bibinfo {author} {\bibfnamefont {G.}~\bibnamefont {Hilt}},
  \bibinfo {author} {\bibfnamefont {X.}~\bibnamefont {Shao}}, \bibinfo {author}
  {\bibfnamefont {J.~M.}\ \bibnamefont {Gottfried}}, \ and\ \bibinfo {author}
  {\bibfnamefont {K.}~\bibnamefont {Wu}},\ }\bibfield  {title} {\enquote
  {\bibinfo {title} {Assembling molecular sierpi{\'{n}}ski triangle fractals}},
  }\href {\doibase 10.1038/nchem.2211} {\bibfield  {journal} {\bibinfo
  {journal} {Nat. Chem.}\ }\textbf {\bibinfo {volume} {7}},\ \bibinfo
  {pages} {389} (\bibinfo {year} {2015})}\BibitemShut {NoStop}%
\bibitem [{\citenamefont {Kempkes}\ \emph {et~al.}(2019)\citenamefont
  {Kempkes}, \citenamefont {Slot}, \citenamefont {Freeney}, \citenamefont
  {Zevenhuizen}, \citenamefont {Vanmaekelbergh}, \citenamefont {Swart},\ and\
  \citenamefont {Smith}}]{Kempkes19nature}%
  \BibitemOpen
  \bibfield  {author} {\bibinfo {author} {\bibfnamefont {S.~N.}\ \bibnamefont
  {Kempkes}}, \bibinfo {author} {\bibfnamefont {M.~R.}\ \bibnamefont {Slot}},
  \bibinfo {author} {\bibfnamefont {S.~E.}\ \bibnamefont {Freeney}}, \bibinfo
  {author} {\bibfnamefont {S.~J.~M.}\ \bibnamefont {Zevenhuizen}}, \bibinfo
  {author} {\bibfnamefont {D.}~\bibnamefont {Vanmaekelbergh}}, \bibinfo
  {author} {\bibfnamefont {I.}~\bibnamefont {Swart}}, \ and\ \bibinfo {author}
  {\bibfnamefont {C.~M.}\ \bibnamefont {Smith}},\ }\bibfield  {title} {\enquote
  {\bibinfo {title} {Design and characterization of electrons in a fractal
  geometry}}, }\href {\doibase 10.1038/s41567-018-0328-0} {\bibfield  {journal}
  {\bibinfo  {journal} {Nat. Phys.}\ }\textbf {\bibinfo {volume} {15}},\
  \bibinfo {pages} {127} (\bibinfo {year} {2019})}\BibitemShut {NoStop}%
\bibitem [{\citenamefont {Liu}\ \emph {et~al.}(2021)\citenamefont {Liu},
  \citenamefont {Zhou}, \citenamefont {Wang}, \citenamefont {Yin},
  \citenamefont {Li}, \citenamefont {Huang}, \citenamefont {Guan},
  \citenamefont {Li}, \citenamefont {Wang}, \citenamefont {Zheng},
  \citenamefont {Liu}, \citenamefont {Han}, \citenamefont {Evans},
  \citenamefont {Liu},\ and\ \citenamefont {Jia}}]{LiuC21prl}%
  \BibitemOpen
  \bibfield  {author} {\bibinfo {author} {\bibfnamefont {C.}~\bibnamefont
  {Liu}}, \bibinfo {author} {\bibfnamefont {Y.}~\bibnamefont {Zhou}}, \bibinfo
  {author} {\bibfnamefont {G.}~\bibnamefont {Wang}}, \bibinfo {author}
  {\bibfnamefont {Y.}~\bibnamefont {Yin}}, \bibinfo {author} {\bibfnamefont
  {C.}~\bibnamefont {Li}}, \bibinfo {author} {\bibfnamefont {H.}~\bibnamefont
  {Huang}},  \emph {et~al.},\ }\bibfield  {title} {\enquote {\bibinfo {title}
  {Sierpi\ifmmode \acute{n}\else \'{n}\fi{}ski structure and electronic
  topology in bi thin films on insb(111)b surfaces}}, }\href {\doibase
  10.1103/PhysRevLett.126.176102} {\bibfield  {journal} {\bibinfo  {journal}
  {Phys. Rev. Lett.}\ }\textbf {\bibinfo {volume} {126}},\ \bibinfo {pages}
  {176102} (\bibinfo {year} {2021})}\BibitemShut {NoStop}%
\bibitem [{\citenamefont {Canyellas}\ \emph {et~al.}(2024)\citenamefont
  {Canyellas}, \citenamefont {Liu}, \citenamefont {Arouca}, \citenamefont
  {Eek}, \citenamefont {Wang}, \citenamefont {Yin}, \citenamefont {Guan}, \citenamefont {Li},\citenamefont {Wang}, \citenamefont {Zheng}, \citenamefont {Liu}, \citenamefont {Jia},\ and\
  \citenamefont {Smith}}]{Canyellas24nature}%
  \BibitemOpen
  \bibfield  {author} {\bibinfo {author} {\bibfnamefont {R.}\ \bibnamefont
  {Canyellas}}, \bibinfo {author} {\bibfnamefont {C.}\ \bibnamefont {Liu}},
  \bibinfo {author} {\bibfnamefont {R.}\ \bibnamefont {Arouca}}, \bibinfo
  {author} {\bibfnamefont {L.}\ \bibnamefont {Eek}}, \bibinfo
  {author} {\bibfnamefont {G.}~\bibnamefont {Wang}}, \bibinfo
  {author} {\bibfnamefont {Y.}~\bibnamefont {Yin}}, \bibinfo
  {author} {\bibfnamefont {D.}~\bibnamefont {Guan}}, \bibinfo
  {author} {\bibfnamefont {Y.}~\bibnamefont {Li}}, \bibinfo
  {author} {\bibfnamefont {S.}~\bibnamefont {Wang}}, \bibinfo
  {author} {\bibfnamefont {H.}~\bibnamefont {Zheng}}, \bibinfo
  {author} {\bibfnamefont {C.}~\bibnamefont {Liu}}, \bibinfo
  {author} {\bibfnamefont {J.}~\bibnamefont {Jia}}, \ and\ \bibinfo {author}
  {\bibfnamefont {C.~M.}\ \bibnamefont {Smith}},\ }\bibfield  {title} {\enquote
  {\bibinfo {title} {Topological edge and corner states in bismuth fractal nanostructures}}}, \href {\doibase 10.34133/2021/5608038} {\bibfield  {journal} {\bibinfo  {journal} {Nat. Phys.}\ }\textbf {\bibinfo {volume} {2024}} (\bibinfo {year} {2024}),\ 10.1038/s41567-024-02551-8}%
\bibitem [{\citenamefont {Lee}(2016)}]{Lee16prl}%
  \BibitemOpen
  \bibfield  {author} {\bibinfo {author} {\bibfnamefont {T.~E.}\ \bibnamefont
  {Lee}},\ }\bibfield  {title} {\enquote {\bibinfo {title} {Anomalous edge
  state in a non-hermitian lattice}}, }\href {\doibase
  10.1103/PhysRevLett.116.133903} {\bibfield  {journal} {\bibinfo  {journal}
  {Phys. Rev. Lett.}\ }\textbf {\bibinfo {volume} {116}},\ \bibinfo {pages}
  {133903} (\bibinfo {year} {2016})}\BibitemShut {NoStop}%
\bibitem [{\citenamefont {Yao}\ and\ \citenamefont {Wang}(2018)}]{Yao18prl}%
  \BibitemOpen
  \bibfield  {author} {\bibinfo {author} {\bibfnamefont {S.}~\bibnamefont
  {Yao}}\ and\ \bibinfo {author} {\bibfnamefont {Z.}~\bibnamefont {Wang}},\
  }\bibfield  {title} {\enquote {\bibinfo {title} {Edge states and topological
  invariants of non-hermitian systems}}, }\href {\doibase
  10.1103/PhysRevLett.121.086803} {\bibfield  {journal} {\bibinfo  {journal}
  {Phys. Rev. Lett.}\ }\textbf {\bibinfo {volume} {121}},\ \bibinfo {pages}
  {086803} (\bibinfo {year} {2018})}\BibitemShut {NoStop}%
\bibitem [{\citenamefont {Kunst}\ \emph {et~al.}(2018)\citenamefont {Kunst},
  \citenamefont {Edvardsson}, \citenamefont {Budich},\ and\ \citenamefont
  {Bergholtz}}]{Kunst18prl}%
  \BibitemOpen
  \bibfield  {author} {\bibinfo {author} {\bibfnamefont {F.~K.}\ \bibnamefont
  {Kunst}}, \bibinfo {author} {\bibfnamefont {E.}~\bibnamefont {Edvardsson}},
  \bibinfo {author} {\bibfnamefont {J.~C.}\ \bibnamefont {Budich}}, \ and\
  \bibinfo {author} {\bibfnamefont {E.~J.}\ \bibnamefont {Bergholtz}},\
  }\bibfield  {title} {\enquote {\bibinfo {title} {Biorthogonal bulk-boundary
  correspondence in non-hermitian systems}}, }\href {\doibase
  10.1103/PhysRevLett.121.026808} {\bibfield  {journal} {\bibinfo  {journal}
  {Phys. Rev. Lett.}\ }\textbf {\bibinfo {volume} {121}},\ \bibinfo {pages}
  {026808} (\bibinfo {year} {2018})}\BibitemShut {NoStop}%
\bibitem [{\citenamefont {Gong}\ \emph {et~al.}(2018)\citenamefont {Gong},
  \citenamefont {Ashida}, \citenamefont {Kawabata}, \citenamefont {Takasan},
  \citenamefont {Higashikawa},\ and\ \citenamefont {Ueda}}]{GonZP18prx}%
  \BibitemOpen
  \bibfield  {author} {\bibinfo {author} {\bibfnamefont {Z.}~\bibnamefont
  {Gong}}, \bibinfo {author} {\bibfnamefont {Y.}~\bibnamefont {Ashida}},
  \bibinfo {author} {\bibfnamefont {K.}~\bibnamefont {Kawabata}}, \bibinfo
  {author} {\bibfnamefont {K.}~\bibnamefont {Takasan}}, \bibinfo {author}
  {\bibfnamefont {S.}~\bibnamefont {Higashikawa}}, \ and\ \bibinfo {author}
  {\bibfnamefont {M.}~\bibnamefont {Ueda}},\ }\bibfield  {title} {\enquote
  {\bibinfo {title} {Topological phases of non-hermitian systems}}, }\href
  {\doibase 10.1103/PhysRevX.8.031079} {\bibfield  {journal} {\bibinfo
  {journal} {Phys. Rev. X}\ }\textbf {\bibinfo {volume} {8}},\ \bibinfo {pages}
  {031079} (\bibinfo {year} {2018})}\BibitemShut {NoStop}%
\bibitem [{\citenamefont {Leykam}\ \emph {et~al.}(2017)\citenamefont {Leykam},
  \citenamefont {Bliokh}, \citenamefont {Huang}, \citenamefont {Chong},\ and\
  \citenamefont {Nori}}]{Leykam17prl}%
  \BibitemOpen
  \bibfield  {author} {\bibinfo {author} {\bibfnamefont {D.}~\bibnamefont
  {Leykam}}, \bibinfo {author} {\bibfnamefont {K.~Y.}\ \bibnamefont {Bliokh}},
  \bibinfo {author} {\bibfnamefont {C.}~\bibnamefont {Huang}}, \bibinfo
  {author} {\bibfnamefont {Y.~D.}\ \bibnamefont {Chong}}, \ and\ \bibinfo
  {author} {\bibfnamefont {F.}~\bibnamefont {Nori}},\ }\bibfield  {title}
  {\enquote {\bibinfo {title} {Edge modes, degeneracies, and topological
  numbers in non-Hermitian systems}}, }\href {\doibase
  10.1103/PhysRevLett.118.040401} {\bibfield  {journal} {\bibinfo  {journal}
  {Phys. Rev. Lett.}\ }\textbf {\bibinfo {volume} {118}},\ \bibinfo {pages}
  {040401} (\bibinfo {year} {2017})}\BibitemShut {NoStop}%
\bibitem [{\citenamefont {Lee}\ \emph {et~al.}(2020)\citenamefont {Lee},
 \ and\ \citenamefont {Thomale}}]{LeeCH19prb}%
  \BibitemOpen
  \bibfield  {author} {\bibinfo {author} {\bibfnamefont {C. H}~\bibnamefont
  {Lee}}, \ and\
  \bibinfo {author} {\bibfnamefont {R.}~\bibnamefont {Thomale}},\ }\bibfield
  {title} {\enquote {\bibinfo {title} {Anatomy of skin modes and topology in non-Hermitian systems}}, }\href {\doibase 10.1103/PhysRevB.99.201103} {\bibfield  {journal} {\bibinfo  {journal}
  {Phys. Rev. B}\ }\textbf {\bibinfo {volume} {99}},\ \bibinfo {pages}
  {201103(R)} (\bibinfo {year} {2019})}\BibitemShut {NoStop}%
\bibitem [{\citenamefont {Zhang}\ \emph {et~al.}(2020)\citenamefont {Zhang},
  \citenamefont {Yang},\ and\ \citenamefont {Fang}}]{ZhangK20prl}%
  \BibitemOpen
  \bibfield  {author} {\bibinfo {author} {\bibfnamefont {K.}~\bibnamefont
  {Zhang}}, \bibinfo {author} {\bibfnamefont {Z.}~\bibnamefont {Yang}}, \ and\
  \bibinfo {author} {\bibfnamefont {C.}~\bibnamefont {Fang}},\ }\bibfield
  {title} {\enquote {\bibinfo {title} {Correspondence between winding numbers
  and skin modes in non-hermitian systems}}, }\href {\doibase
  10.1103/PhysRevLett.125.126402} {\bibfield  {journal} {\bibinfo  {journal}
  {Phys. Rev. Lett.}\ }\textbf {\bibinfo {volume} {125}},\ \bibinfo {pages}
  {126402} (\bibinfo {year} {2020})}\BibitemShut {NoStop}%
\bibitem [{\citenamefont {Shen}\ \emph {et~al.}(2018)\citenamefont {Shen},
  \citenamefont {Zhen},\ and\ \citenamefont {Fu}}]{ShenH18prl}%
  \BibitemOpen
  \bibfield  {author} {\bibinfo {author} {\bibfnamefont {H.}~\bibnamefont
  {Shen}}, \bibinfo {author} {\bibfnamefont {B.}~\bibnamefont {Zhen}}, \ and\
  \bibinfo {author} {\bibfnamefont {L.}~\bibnamefont {Fu}},\ }\bibfield
  {title} {\enquote {\bibinfo {title} {Topological band theory for
  non-hermitian hamiltonians}}, }\href {\doibase
  10.1103/PhysRevLett.120.146402} {\bibfield  {journal} {\bibinfo  {journal}
  {Phys. Rev. Lett.}\ }\textbf {\bibinfo {volume} {120}},\ \bibinfo {pages}
  {146402} (\bibinfo {year} {2018})}\BibitemShut {NoStop}%
\bibitem [{\citenamefont {Okuma}\ \emph {et~al.}(2020)\citenamefont {Okuma},
  \citenamefont {Kawabata}, \citenamefont {Shiozaki},\ and\ \citenamefont
  {Sato}}]{Okuma20prl}%
  \BibitemOpen
  \bibfield  {author} {\bibinfo {author} {\bibfnamefont {N.}~\bibnamefont
  {Okuma}}, \bibinfo {author} {\bibfnamefont {K.}~\bibnamefont {Kawabata}},
  \bibinfo {author} {\bibfnamefont {K.}~\bibnamefont {Shiozaki}}, \ and\
  \bibinfo {author} {\bibfnamefont {M.}~\bibnamefont {Sato}},\ }\bibfield
  {title} {\enquote {\bibinfo {title} {Topological origin of non-Hermitian skin
  effects}}, }\href {\doibase 10.1103/PhysRevLett.124.086801} {\bibfield
  {journal} {\bibinfo  {journal} {Phys. Rev. Lett.}\ }\textbf {\bibinfo
  {volume} {124}},\ \bibinfo {pages} {086801} (\bibinfo {year}
  {2020})}\BibitemShut {NoStop}%
\bibitem [{\citenamefont {Kawabata}\ \emph {et~al.}(2019)\citenamefont
  {Kawabata}, \citenamefont {Bessho},\ and\ \citenamefont
  {Sato}}]{Kawabata19PRL}%
  \BibitemOpen
  \bibfield  {author} {\bibinfo {author} {\bibfnamefont {K.}~\bibnamefont
  {Kawabata}}, \bibinfo {author} {\bibfnamefont {T.}~\bibnamefont {Bessho}}, \
  and\ \bibinfo {author} {\bibfnamefont {M.}~\bibnamefont {Sato}},\ }\bibfield
  {title} {\enquote {\bibinfo {title} {Classification of exceptional points and
  non-Hermitian topological semimetals}}, }\href {\doibase
  10.1103/PhysRevLett.123.066405} {\bibfield  {journal} {\bibinfo  {journal}
  {Phys. Rev. Lett.}\ }\textbf {\bibinfo {volume} {123}},\ \bibinfo {pages}
  {066405} (\bibinfo {year} {2019})}\BibitemShut {NoStop}%
\bibitem [{\citenamefont {Yokomizo}\ and\ \citenamefont
  {Murakami}(2019)}]{Yokomizo19prl}%
  \BibitemOpen
  \bibfield  {author} {\bibinfo {author} {\bibfnamefont {K.}~\bibnamefont
  {Yokomizo}}\ and\ \bibinfo {author} {\bibfnamefont {S.}~\bibnamefont
  {Murakami}},\ }\bibfield  {title} {\enquote {\bibinfo {title} {Non-bloch band
  theory of non-Hermitian systems}}, }\href {\doibase
  10.1103/PhysRevLett.123.066404} {\bibfield  {journal} {\bibinfo  {journal}
  {Phys. Rev. Lett.}\ }\textbf {\bibinfo {volume} {123}},\ \bibinfo {pages}
  {066404} (\bibinfo {year} {2019})}\BibitemShut {NoStop}%
\bibitem [{\citenamefont {Zhou}\ and\ \citenamefont {Lee}(2019)}]{Zhou19prb}%
  \BibitemOpen
  \bibfield  {author} {\bibinfo {author} {\bibfnamefont {H.}~\bibnamefont
  {Zhou}}\ and\ \bibinfo {author} {\bibfnamefont {J.~Y.}\ \bibnamefont {Lee}},\
  }\bibfield  {title} {\enquote {\bibinfo {title} {Periodic table for
  topological bands with non-Hermitian symmetries}}, }\href {\doibase
  10.1103/PhysRevB.99.235112} {\bibfield  {journal} {\bibinfo  {journal} {Phys.
  Rev. B}\ }\textbf {\bibinfo {volume} {99}},\ \bibinfo {pages} {235112}
  (\bibinfo {year} {2019})}\BibitemShut {NoStop}%
\bibitem [{\citenamefont {Borgnia}\ \emph {et~al.}(2020)\citenamefont
  {Borgnia}, \citenamefont {Kruchkov},\ and\ \citenamefont
  {Slager}}]{Borgnia20prl}%
  \BibitemOpen
  \bibfield  {author} {\bibinfo {author} {\bibfnamefont {D.~S.}\ \bibnamefont
  {Borgnia}}, \bibinfo {author} {\bibfnamefont {A.~J.}\ \bibnamefont
  {Kruchkov}}, \ and\ \bibinfo {author} {\bibfnamefont {R.-J.}\ \bibnamefont
  {Slager}},\ }\bibfield  {title} {\enquote {\bibinfo {title} {Non-hermitian
  boundary modes and topology}}, }\href {\doibase
  10.1103/PhysRevLett.124.056802} {\bibfield  {journal} {\bibinfo  {journal}
  {Phys. Rev. Lett.}\ }\textbf {\bibinfo {volume} {124}},\ \bibinfo {pages}
  {056802} (\bibinfo {year} {2020})}\BibitemShut {NoStop}%
\bibitem [{\citenamefont {Budich}\ and\ \citenamefont
  {Bergholtz}(2020)}]{Budich20prl}%
  \BibitemOpen
  \bibfield  {author} {\bibinfo {author} {\bibfnamefont {J.~C.}\ \bibnamefont
  {Budich}}\ and\ \bibinfo {author} {\bibfnamefont {E.~J.}\ \bibnamefont
  {Bergholtz}},\ }\bibfield  {title} {\enquote {\bibinfo {title} {Non-Hermitian
  topological sensors}}, }\href {\doibase 10.1103/PhysRevLett.125.180403}
  {\bibfield  {journal} {\bibinfo  {journal} {Phys. Rev. Lett.}\ }\textbf
  {\bibinfo {volume} {125}},\ \bibinfo {pages} {180403} (\bibinfo {year}
  {2020})}\BibitemShut {NoStop}%
\bibitem [{\citenamefont {Bergholtz}\ \emph {et~al.}(2021)\citenamefont
  {Bergholtz}, \citenamefont {Budich},\ and\ \citenamefont
  {Kunst}}]{Bergholtz21rmp}%
  \BibitemOpen
  \bibfield  {author} {\bibinfo {author} {\bibfnamefont {E.~J.}\ \bibnamefont
  {Bergholtz}}, \bibinfo {author} {\bibfnamefont {J.~C.}\ \bibnamefont
  {Budich}}, \ and\ \bibinfo {author} {\bibfnamefont {F.~K.}\ \bibnamefont
  {Kunst}},\ }\bibfield  {title} {\enquote {\bibinfo {title} {Exceptional
  topology of non-Hermitian systems}}, }\href {\doibase
  10.1103/RevModPhys.93.015005} {\bibfield  {journal} {\bibinfo  {journal}
  {Rev. Mod. Phys.}\ }\textbf {\bibinfo {volume} {93}},\ \bibinfo {pages}
  {015005} (\bibinfo {year} {2021})}\BibitemShut {NoStop}%
  \bibitem [{\citenamefont {Sun}\ \emph {et~al.}(2023)\citenamefont {Sun}, \citenamefont {Li}, \citenamefont {Feng},\ and\ \citenamefont {Guo}}]{PhysRevB.108.075122}%
  \BibitemOpen
  \bibfield  {author} {\bibinfo {author} {\bibfnamefont {J.}~\bibnamefont {Sun}}, \bibinfo {author} {\bibfnamefont {C.-A.}\ \bibnamefont {Li}}, \bibinfo {author} {\bibfnamefont {S.}~\bibnamefont {Feng}}, \ and\ \bibinfo {author} {\bibfnamefont {H.}~\bibnamefont {Guo}},\ }\bibfield  {title} {\enquote {\bibinfo {title} {Hybrid higher-order skin-topological effect in hyperbolic lattices}}, }\href {\doibase 10.1103/PhysRevB.108.075122} {\bibfield  {journal} {\bibinfo  {journal} {Phys. Rev. B}\ }\textbf {\bibinfo {volume} {108}},\ \bibinfo {pages} {075122} (\bibinfo {year} {2023})}\BibitemShut {NoStop}%
\bibitem [{\citenamefont {Li}\ \emph {et~al.}(2023)\citenamefont {Li}, \citenamefont {Trauzettel}, \citenamefont {Neupert},\ and\ \citenamefont {Zhang}}]{PRL-LichangAn}%
  \BibitemOpen
  \bibfield  {author} {\bibinfo {author} {\bibfnamefont {C.-A.}\ \bibnamefont {Li}}, \bibinfo {author} {\bibfnamefont {B.}~\bibnamefont {Trauzettel}}, \bibinfo {author} {\bibfnamefont {T.}~\bibnamefont {Neupert}}, \ and\ \bibinfo {author} {\bibfnamefont {S.-B.}\ \bibnamefont {Zhang}},\ }\bibfield  {title} {\enquote {\bibinfo {title} {Enhancement of second-order non-hermitian skin effect by magnetic fields}}, }\href {\doibase 10.1103/PhysRevLett.131.116601} {\bibfield  {journal} {\bibinfo  {journal} {Phys. Rev. Lett.}\ }\textbf {\bibinfo {volume} {131}},\ \bibinfo {pages} {116601} (\bibinfo {year} {2023})}\BibitemShut {NoStop}%
\bibitem [{\citenamefont {Hatano}\ and\ \citenamefont {Nelson}(1996)}]{PhysRevLett.77.570}%
  \BibitemOpen
  \bibfield  {author} {\bibinfo {author} {\bibfnamefont {N.}~\bibnamefont {Hatano}}\ and\ \bibinfo {author} {\bibfnamefont {D.~R.}\ \bibnamefont {Nelson}},\ }\bibfield  {title} {\enquote {\bibinfo {title} {Localization transitions in non-Hermitian quantum mechanics}}, }\href {\doibase 10.1103/PhysRevLett.77.570} {\bibfield  {journal} {\bibinfo  {journal} {Phys. Rev. Lett.}\ }\textbf {\bibinfo {volume} {77}},\ \bibinfo {pages} {570} (\bibinfo {year} {1996})}\BibitemShut {NoStop}%
\bibitem [{\citenamefont {Hatano}\ and\ \citenamefont {Nelson}(1997)}]{PhysRevB.56.8651}%
  \BibitemOpen
  \bibfield  {author} {\bibinfo {author} {\bibfnamefont {N.}~\bibnamefont {Hatano}}\ and\ \bibinfo {author} {\bibfnamefont {D.~R.}\ \bibnamefont {Nelson}},\ }\bibfield  {title} {\enquote {\bibinfo {title} {Vortex pinning and non-hermitian quantum mechanics}}, }\href {\doibase 10.1103/PhysRevB.56.8651} {\bibfield  {journal} {\bibinfo  {journal} {Phys. Rev. B}\ }\textbf {\bibinfo {volume} {56}},\ \bibinfo {pages} {8651} (\bibinfo {year} {1997})}\BibitemShut {NoStop}%
\bibitem [{SM()}]{SM}%
  \BibitemOpen
  \href@noop {} {}\bibinfo {note} {See Supplemental Material for more details on (S1) Solutions of coupled Hatano-Nelson models on the tree lattice; (S2) Mandelbrot matrix and its characteristic polynomials; (S3) Generalizations of the tree lattice; (S4) Connection of NHQFs to Hofstadter problems with an imaginary magnetic field; (S5) Solutions of NHQFs on a Sierpi$\acute{\mathrm{n}}$ski gasket, which includes the reference \cite{afraceigenvector}. }\BibitemShut {Stop}%
  \bibitem [{Note1()}]{Note1}%
  \BibitemOpen
  \href@noop {} {}\bibinfo {note} {Note that the coupled Hatatno-Nelson models are not at an exceptional point even at $\gamma=t$, which is clear from the Hamiltonian $H_2$ not being a Jordan canonical form.}\BibitemShut {Stop}%
\bibitem [{\citenamefont {Neil J.~Calkin}\ and\ \citenamefont {Lawrence}(2022)}]{afraceigenvector}%
  \BibitemOpen
  \bibfield  {author} {\bibinfo {author} {\bibnamefont {N. J.~Calkin}, \bibfnamefont {E. Y. S.~Chan}}, \bibnamefont {R. M. Corless}, \bibnamefont {D. J.~Jeffery}, \ and\ \bibinfo {author} {\bibfnamefont {P.~W.}\ \bibnamefont {Lawrence}},\ }\bibfield  {title} {\enquote {\bibinfo {title} {A fractal eigenvector}}, }\href {\doibase 10.1080/00029890.2022.2059311} {\bibfield  {journal} {\bibinfo  {journal} {Amer. Math. Monthly}\ }\textbf {\bibinfo {volume} {129}},\ \bibinfo {pages} {503} (\bibinfo {year} {2022})} \BibitemShut {NoStop}%
\bibitem [{\citenamefont {Chan}\ and\ \citenamefont {Corless}(2017)}]{Chan2017ANK}%
  \BibitemOpen
  \bibfield  {author} {\bibinfo {author} {\bibfnamefont {E.~Y.~S.}\ \bibnamefont {Chan}}\ and\ \bibinfo {author} {\bibfnamefont {R.~M.}\ \bibnamefont {Corless}},\ }\bibfield  {title} {\enquote {\bibinfo {title} {A new kind of companion matrix}}, }\href {https://api.semanticscholar.org/CorpusID:126086320} {\bibfield  {journal} {\bibinfo  {journal} {Electron. J. Linear Algebra}\ }\textbf {\bibinfo {volume} {32}},\ \bibinfo {pages} {335} (\bibinfo {year} {2017})}\BibitemShut {NoStop}%
\bibitem [{\citenamefont {Peitgen}\ \emph {et~al.}(2004)\citenamefont {Peitgen}, \citenamefont {J{\"u}rgens}, \citenamefont {Saupe},\ and\ \citenamefont {Feigenbaum}}]{chaosandfractals}%
  \BibitemOpen
  \bibfield  {author} {\bibinfo {author} {\bibfnamefont {H.-O.}\ \bibnamefont {Peitgen}}, \bibinfo {author} {\bibfnamefont {H.}~\bibnamefont {J{\"u}rgens}}, \bibinfo {author} {\bibfnamefont {D.}~\bibnamefont {Saupe}}, \ and\ \bibinfo {author} {\bibfnamefont {M.~J.}\ \bibnamefont {Feigenbaum}},\ }\href@noop {} {\emph {\bibinfo {title} {Chaos and fractals: new frontiers of science}}},\ Vol.\ \bibinfo {volume} {106}\ (\bibinfo  {publisher} {Springer},\ \bibinfo {year} {2004})\BibitemShut {NoStop}%
\bibitem [{\citenamefont {Chan}(2016)}]{Mandelbrot-like}%
  \BibitemOpen
  \bibfield  {author} {\bibinfo {author} {\bibfnamefont {E.~Y.}\ \bibnamefont {Chan}},\ }\emph {\bibinfo {title} {A comparison of solution methods for Mandelbrot-like polynomials}},\ \href@noop {} {Ph.D. thesis},\ \bibinfo  {school} {The University of Western Ontario (Canada)} (\bibinfo {year} {2016})\BibitemShut {NoStop}%
\bibitem [{\citenamefont {Mandelbrot}(1980)}]{mandelbrot1980fractal}%
  \BibitemOpen
  \bibfield  {author} {\bibinfo {author} {\bibfnamefont {B.~B.}\ \bibnamefont {Mandelbrot}},\ }\bibfield  {title} {\enquote {\bibinfo {title} {Fractal aspects of the iteration of z$\rightarrow \lambda$z (1-z) for complex $\lambda$ and z}}, }\href@noop {} {\bibfield  {journal} {\bibinfo  {journal} {Annals of the New York Academy of Sciences}\ }\textbf {\bibinfo {volume} {357}},\ \bibinfo {pages} {249} (\bibinfo {year} {1980})}\BibitemShut {NoStop}%
\bibitem [{\citenamefont {Devaney}(1999)}]{devaney1999mandelbrot}%
  \BibitemOpen
  \bibfield  {author} {\bibinfo {author} {\bibfnamefont {R.~L.}\ \bibnamefont {Devaney}},\ }\bibfield  {title} {\enquote {\bibinfo {title} {The Mandelbrot set, the farey tree, and the fibonacci sequence}}, }\href@noop {} {\bibfield  {journal} {\bibinfo  {journal} {Amer. Math. Monthly}\ }\textbf {\bibinfo {volume} {106}},\ \bibinfo {pages} {289} (\bibinfo {year} {1999})}\BibitemShut {NoStop}%
\bibitem [{\citenamefont {Rochon}(2000)}]{rochon2000generalized}%
  \BibitemOpen
  \bibfield  {author} {\bibinfo {author} {\bibfnamefont {D.}~\bibnamefont {Rochon}},\ }\bibfield  {title} {\enquote {\bibinfo {title} {A generalized Mandelbrot set for bicomplex numbers}}, }\href@noop {} {\bibfield  {journal} {\bibinfo  {journal} {Fractals}\ }\textbf {\bibinfo {volume} {8}},\ \bibinfo {pages} {355} (\bibinfo {year} {2000})}\BibitemShut {NoStop}%
\bibitem [{\citenamefont {Lei}(1990)}]{lei1990similarity}%
  \BibitemOpen
  \bibfield  {author} {\bibinfo {author} {\bibfnamefont {T.}~\bibnamefont {Lei}},\ }\bibfield  {title} {\enquote {\bibinfo {title} {Similarity between the Mandelbrot set and julia sets}}, }\href@noop {} {\bibfield  {journal} {\bibinfo  {journal} {Commun. Math. Phys.}\ }\textbf {\bibinfo {volume} {134}},\ \bibinfo {pages} {587} (\bibinfo {year} {1990})}\BibitemShut {NoStop}%
 \bibitem [{Note1()}]{Note3}%
  \BibitemOpen
  \href@noop {} {}\bibinfo {note} {The Hofstadter fractals result from the incommensurability of the magnetic length $\ell_B=\sqrt{\hbar/eB}$ with a magnetic field $B$ and the lattice constant.}\BibitemShut {Stop}%
\bibitem [{\citenamefont {Petrova}\ \emph {et~al.}(2024)\citenamefont {Petrova}, \citenamefont {Tiunov}, \citenamefont {Ba\~nuls},\ and\ \citenamefont {Fedorov}}]{PhysRevLett.132.050401}%
  \BibitemOpen
  \bibfield  {author} {\bibinfo {author} {\bibfnamefont {E.~V.}\ \bibnamefont {Petrova}}, \bibinfo {author} {\bibfnamefont {E.~S.}\ \bibnamefont {Tiunov}}, \bibinfo {author} {\bibfnamefont {M.~C.}\ \bibnamefont {Ba\~nuls}}, \ and\ \bibinfo {author} {\bibfnamefont {A.~K.}\ \bibnamefont {Fedorov}},\ }\bibfield  {title} {\enquote {\bibinfo {title} {Fractal states of the schwinger model}}, }\href {\doibase 10.1103/PhysRevLett.132.050401} {\bibfield  {journal} {\bibinfo  {journal} {Phys. Rev. Lett.}\ }\textbf {\bibinfo {volume} {132}},\ \bibinfo {pages} {050401} (\bibinfo {year} {2024})}\BibitemShut {NoStop}%
\bibitem [{\citenamefont {Chen}\ \emph {et~al.}(2024)\citenamefont {Chen}, \citenamefont {Lou}, \citenamefont {Hu},\ and\ \citenamefont {Lim}}]{chen2024fractal}%
  \BibitemOpen
  \bibfield  {author} {\bibinfo {author} {\bibfnamefont {Z.-G.}\ \bibnamefont {Chen}}, \bibinfo {author} {\bibfnamefont {C.}~\bibnamefont {Lou}}, \bibinfo {author} {\bibfnamefont {K.}~\bibnamefont {Hu}}, \ and\ \bibinfo {author} {\bibfnamefont {L.-K.}\ \bibnamefont {Lim}},\ }\href@noop {} {\enquote {\bibinfo {title} {Fractal surface states in three-dimensional topological quasicrystals}}, } (\bibinfo {year} {2024}),\ \Eprint {http://arxiv.org/abs/2401.11497} {arXiv:2401.11497 [cond-mat.mes-hall]} \BibitemShut {NoStop}%
\bibitem [{\citenamefont {Wan}\ \emph {et~al.}(2024)\citenamefont {Wan}, \citenamefont {Gao},\ and\ \citenamefont {Shi}}]{wan2024fractal}%
  \BibitemOpen
  \bibfield  {author} {\bibinfo {author} {\bibfnamefont {X.-T.}\ \bibnamefont {Wan}}, \bibinfo {author} {\bibfnamefont {C.}~\bibnamefont {Gao}}, \ and\ \bibinfo {author} {\bibfnamefont {Z.-Y.}\ \bibnamefont {Shi}},\ }\href@noop {} {\enquote {\bibinfo {title} {Fractal spectrum in twisted bilayer optical lattice}}, } (\bibinfo {year} {2024}),\ \Eprint {http://arxiv.org/abs/2404.08211} {arXiv:2404.08211 [cond-mat.mes-hall]} \BibitemShut {NoStop}%
\bibitem [{\citenamefont {Rodríguez-Laguna}\ \emph {et~al.}(2012)\citenamefont {Rodríguez-Laguna}, \citenamefont {Migdał}, \citenamefont {Berganza}, \citenamefont {Lewenstein},\ and\ \citenamefont {Sierra}}]{Rod2012}%
  \BibitemOpen
  \bibfield  {author} {\bibinfo {author} {\bibfnamefont {J.}~\bibnamefont {Rodr\'iguez-Laguna}}, \bibinfo {author} {\bibfnamefont {P.}~\bibnamefont {Miguel}}, \bibinfo {author} {\bibfnamefont {M.~I.}\ \bibnamefont {Berganza}}, \bibinfo {author} {\bibfnamefont {M.}~\bibnamefont {Lewenstein}}, \ and\ \bibinfo {author} {\bibfnamefont {G.}~\bibnamefont {Sierra}},\ }\bibfield  {title} {\enquote {\bibinfo {title} {Qubism: self-similar visualization of many-body wavefunctions}}, }\href {\doibase 10.1088/1367-2630/14/5/053028} {\bibfield  {journal} {\bibinfo  {journal} {New J. Phys.}\ }\textbf {\bibinfo {volume} {14}},\ \bibinfo {pages} {053028} (\bibinfo {year} {2012})}\BibitemShut {NoStop}%
\bibitem [{\citenamefont {Xiujuan~Zhang}\ and\ \citenamefont {Chen}(2022)}]{Advance-review}%
  \BibitemOpen
  \bibfield  {author} {\bibinfo {author} {\bibfnamefont {M.-H.~Lu},\ \bibnamefont {X.~Zhang}, \bibfnamefont {T.~Zhang}}\ and\ \bibinfo {author} {\bibfnamefont {Y.-F.}\ \bibnamefont {Chen}},\ }\bibfield  {title} {\enquote {\bibinfo {title} {A review on non-hermitian skin effect}}, }\href {\doibase 10.1080/23746149.2022.2109431} {\bibfield  {journal} {\bibinfo  {journal} {Adv. Phys.: X}\ }\textbf {\bibinfo {volume} {7}},\ \bibinfo {pages} {2109431} (\bibinfo {year} {2022})} \BibitemShut {NoStop}%
\bibitem [{\citenamefont {Roy}\ \emph {et~al.}(2021)\citenamefont {Roy}, \citenamefont {Mishra}, \citenamefont {Tanatar},\ and\ \citenamefont {Basu}}]{ReentrantLocalization}%
  \BibitemOpen
  \bibfield  {author} {\bibinfo {author} {\bibfnamefont {S.}~\bibnamefont {Roy}}, \bibinfo {author} {\bibfnamefont {T.}~\bibnamefont {Mishra}}, \bibinfo {author} {\bibfnamefont {B.}~\bibnamefont {Tanatar}}, \ and\ \bibinfo {author} {\bibfnamefont {S.}~\bibnamefont {Basu}},\ }\bibfield  {title} {\enquote {\bibinfo {title} {Reentrant localization transition in a quasiperiodic chain}}, }\href {\doibase 10.1103/PhysRevLett.126.106803} {\bibfield  {journal} {\bibinfo  {journal} {Phys. Rev. Lett.}\ }\textbf {\bibinfo {volume} {126}},\ \bibinfo {pages} {106803} (\bibinfo {year} {2021})}\BibitemShut {NoStop}%
\bibitem [{\citenamefont {Li}\ \emph {et~al.}(2017)\citenamefont {Li}, \citenamefont {Li},\ and\ \citenamefont {Das~Sarma}}]{LiXiao}%
  \BibitemOpen
  \bibfield  {author} {\bibinfo {author} {\bibfnamefont {X.}~\bibnamefont {Li}}, \bibinfo {author} {\bibfnamefont {X.}~\bibnamefont {Li}}, \ and\ \bibinfo {author} {\bibfnamefont {S.}~\bibnamefont {Das~Sarma}},\ }\bibfield  {title} {\enquote {\bibinfo {title} {Mobility edges in one-dimensional bichromatic incommensurate potentials}}, }\href {\doibase 10.1103/PhysRevB.96.085119} {\bibfield  {journal} {\bibinfo  {journal} {Phys. Rev. B}\ }\textbf {\bibinfo {volume} {96}},\ \bibinfo {pages} {085119} (\bibinfo {year} {2017})}\BibitemShut {NoStop}%
\bibitem [{\citenamefont {Li}\ \emph {et~al.}(2022{\natexlab{b}})\citenamefont {Li}, \citenamefont {Zhang}, \citenamefont {Budich},\ and\ \citenamefont {Trauzettel}}]{LiChang-An}%
  \BibitemOpen
  \bibfield  {author} {\bibinfo {author} {\bibfnamefont {C.-A.}\ \bibnamefont {Li}}, \bibinfo {author} {\bibfnamefont {S.-B.}\ \bibnamefont {Zhang}}, \bibinfo {author} {\bibfnamefont {J.~C.}\ \bibnamefont {Budich}}, \ and\ \bibinfo {author} {\bibfnamefont {B.}~\bibnamefont {Trauzettel}},\ }\bibfield  {title} {\enquote {\bibinfo {title} {Transition from metal to higher-order topological insulator driven by random flux}}, }\href {\doibase 10.1103/PhysRevB.106.L081410} {\bibfield  {journal} {\bibinfo  {journal} {Phys. Rev. B}\ }\textbf {\bibinfo {volume} {106}},\ \bibinfo {pages} {L081410} (\bibinfo {year} {2022}{\natexlab{b}})}\BibitemShut {NoStop}%
\bibitem [{\citenamefont {Lee}\ \emph {et~al.}(2018)\citenamefont {Lee}, \citenamefont {Imhof}, \citenamefont {Berger}, \citenamefont {Bayer}, \citenamefont {Brehm}, \citenamefont {Molenkamp}, \citenamefont {Kiessling},\ and\ \citenamefont {Thomale}}]{Lee2018}%
  \BibitemOpen
  \bibfield  {author} {\bibinfo {author} {\bibfnamefont {C.~H.}\ \bibnamefont {Lee}}, \bibinfo {author} {\bibfnamefont {S.}~\bibnamefont {Imhof}}, \bibinfo {author} {\bibfnamefont {C.}~\bibnamefont {Berger}}, \bibinfo {author} {\bibfnamefont {F.}~\bibnamefont {Bayer}}, \bibinfo {author} {\bibfnamefont {J.}~\bibnamefont {Brehm}}, \bibinfo {author} {\bibfnamefont {L.~W.}\ \bibnamefont {Molenkamp}}, \bibinfo {author} {\bibfnamefont {T.}~\bibnamefont {Kiessling}}, \ and\ \bibinfo {author} {\bibfnamefont {R.}~\bibnamefont {Thomale}},\ }\bibfield  {title} {\enquote {\bibinfo {title} {Topolectrical circuits}}, }\href {\doibase 10.1038/s42005-018-0035-2} {\bibfield  {journal} {\bibinfo  {journal} {Commun. Phys.}\ }\textbf {\bibinfo {volume} {1}},\ \bibinfo {pages} {39} (\bibinfo {year} {2018})}\BibitemShut {NoStop}%
\bibitem [{\citenamefont {Imhof}\ \emph {et~al.}(2018)\citenamefont {Imhof}, \citenamefont {Berger}, \citenamefont {Bayer}, \citenamefont {Brehm}, \citenamefont {Molenkamp}, \citenamefont {Kiessling}, \citenamefont {Schindler}, \citenamefont {Lee}, \citenamefont {Greiter}, \citenamefont {Neupert},\ and\ \citenamefont {Thomale}}]{Imhof2018}%
  \BibitemOpen
  \bibfield  {author} {\bibinfo {author} {\bibfnamefont {S.}~\bibnamefont {Imhof}}, \bibinfo {author} {\bibfnamefont {C.}~\bibnamefont {Berger}}, \bibinfo {author} {\bibfnamefont {F.}~\bibnamefont {Bayer}}, \bibinfo {author} {\bibfnamefont {J.}~\bibnamefont {Brehm}}, \bibinfo {author} {\bibfnamefont {L.~W.}\ \bibnamefont {Molenkamp}}, \bibinfo {author} {\bibfnamefont {T.}~\bibnamefont {Kiessling}}, \bibinfo {author} {\bibfnamefont {F.}~\bibnamefont {Schindler}}, \bibinfo {author} {\bibfnamefont {C.~H.}\ \bibnamefont {Lee}}, \bibinfo {author} {\bibfnamefont {M.}~\bibnamefont {Greiter}}, \bibinfo {author} {\bibfnamefont {T.}~\bibnamefont {Neupert}},  \emph {et~al.},\ }\bibfield  {title} {\enquote {\bibinfo {title} {Topolectrical-circuit realization of topological corner modes}}, }\href {\doibase 10.1038/s41567-018-0246-1} {\bibfield  {journal} {\bibinfo  {journal} {Nat. Phys.}\ }\textbf {\bibinfo {volume} {14}},\ \bibinfo {pages} {925} (\bibinfo {year} {2018})}\BibitemShut {NoStop}%
\bibitem [{\citenamefont {Zou}\ \emph {et~al.}(2021)\citenamefont {Zou}, \citenamefont {Chen}, \citenamefont {He}, \citenamefont {Bao}, \citenamefont {Lee}, \citenamefont {Sun},\ and\ \citenamefont {Zhang}}]{Zou2021}%
  \BibitemOpen
  \bibfield  {author} {\bibinfo {author} {\bibfnamefont {D.}~\bibnamefont {Zou}}, \bibinfo {author} {\bibfnamefont {T.}~\bibnamefont {Chen}}, \bibinfo {author} {\bibfnamefont {W.}~\bibnamefont {He}}, \bibinfo {author} {\bibfnamefont {J.}~\bibnamefont {Bao}}, \bibinfo {author} {\bibfnamefont {C.~H.}\ \bibnamefont {Lee}}, \bibinfo {author} {\bibfnamefont {H.}~\bibnamefont {Sun}}, \ and\ \bibinfo {author} {\bibfnamefont {X.}~\bibnamefont {Zhang}},\ }\bibfield  {title} {\enquote {\bibinfo {title} {Observation of hybrid higher-order skin-topological effect in non-hermitian topolectrical circuits}}, }\href {\doibase 10.1038/s41467-021-26414-5} {\bibfield  {journal} {\bibinfo  {journal} {Nat. Commun.}\ }\textbf {\bibinfo {volume} {12}},\ \bibinfo {pages} {7201} (\bibinfo {year} {2021})}\BibitemShut {NoStop}%
\bibitem [{\citenamefont {Liu}\ \emph {et~al.}(2021)\citenamefont {Liu}, \citenamefont {Shao}, \citenamefont {Ma}, \citenamefont {Zhang}, \citenamefont {You}, \citenamefont {Wu}, \citenamefont {Xiang}, \citenamefont {Cui},\ and\ \citenamefont {Zhang}}]{ShuoLiu}%
  \BibitemOpen
  \bibfield  {author} {\bibinfo {author} {\bibfnamefont {S.}~\bibnamefont {Liu}}, \bibinfo {author} {\bibfnamefont {R.}~\bibnamefont {Shao}}, \bibinfo {author} {\bibfnamefont {S.}~\bibnamefont {Ma}}, \bibinfo {author} {\bibfnamefont {L.}~\bibnamefont {Zhang}}, \bibinfo {author} {\bibfnamefont {O.}~\bibnamefont {You}}, \bibinfo {author} {\bibfnamefont {H.}~\bibnamefont {Wu}}, \bibinfo {author} {\bibfnamefont {Y.~J.}\ \bibnamefont {Xiang}}, \bibinfo {author} {\bibfnamefont {T.~J.}\ \bibnamefont {Cui}}, \ and\ \bibinfo {author} {\bibfnamefont {S.}~\bibnamefont {Zhang}},\ }\bibfield  {title} {\enquote {\bibinfo {title} {Non-hermitian skin effect in a non-hermitian electrical circuit}}, }\href {\doibase 10.34133/2021/5608038} {\bibfield  {journal} {\bibinfo  {journal} {Research}\ }\textbf {\bibinfo {volume} {2021}} (\bibinfo {year} {2021}),\ 10.34133/2021/5608038}
  \BibitemShut {NoStop}%
\bibitem [{\citenamefont {Xu}\ \emph {et~al.}(2021)\citenamefont {Xu}, \citenamefont {Zhang}, \citenamefont {Luo}, \citenamefont {Yu}, \citenamefont {Li},\ and\ \citenamefont {Zhang}}]{PhysRevB.103.125411}%
  \BibitemOpen
  \bibfield  {author} {\bibinfo {author} {\bibfnamefont {K.}~\bibnamefont {Xu}}, \bibinfo {author} {\bibfnamefont {X.}~\bibnamefont {Zhang}}, \bibinfo {author} {\bibfnamefont {K.}~\bibnamefont {Luo}}, \bibinfo {author} {\bibfnamefont {R.}~\bibnamefont {Yu}}, \bibinfo {author} {\bibfnamefont {D.}~\bibnamefont {Li}}, \ and\ \bibinfo {author} {\bibfnamefont {H.}~\bibnamefont {Zhang}},\ }\bibfield  {title} {\enquote {\bibinfo {title} {Coexistence of topological edge states and skin effects in the non-hermitian su-schrieffer-heeger model with long-range non-reciprocal hopping in topoelectric realizations}}, }\href {\doibase 10.1103/PhysRevB.103.125411} {\bibfield  {journal} {\bibinfo  {journal} {Phys. Rev. B}\ }\textbf {\bibinfo {volume} {103}},\ \bibinfo {pages} {125411} (\bibinfo {year} {2021})}\BibitemShut {NoStop}%
\bibitem [{\citenamefont {Parto}\ \emph {et~al.}(2021)\citenamefont {Parto}, \citenamefont {Liu}, \citenamefont {Bahari}, \citenamefont {Khajavikhan},\ and\ \citenamefont {Christodoulides}}]{Nanophoton}%
  \BibitemOpen
  \bibfield  {author} {\bibinfo {author} {\bibfnamefont {M.}~\bibnamefont {Parto}}, \bibinfo {author} {\bibfnamefont {Y.~G.~N.}\ \bibnamefont {Liu}}, \bibinfo {author} {\bibfnamefont {B.}~\bibnamefont {Bahari}}, \bibinfo {author} {\bibfnamefont {M.}~\bibnamefont {Khajavikhan}}, \ and\ \bibinfo {author} {\bibfnamefont {D.~N.}\ \bibnamefont {Christodoulides}},\ }\bibfield  {title} {\enquote {\bibinfo {title} {Non-hermitian and topological photonics: optics at an exceptional point}}, }\href {\doibase doi:10.1515/nanoph-2020-0434} {\bibfield  {journal} {\bibinfo  {journal} {Nanophotonics}\ }\textbf {\bibinfo {volume} {10}},\ \bibinfo {pages} {403} (\bibinfo {year} {2021})}\BibitemShut {NoStop}%
\bibitem [{\citenamefont {Qian}\ \emph {et~al.}(2024)\citenamefont {Qian}, \citenamefont {Li}, \citenamefont {Zhu}, \citenamefont {You},\ and\ \citenamefont {Wang}}]{PhysRevLett.132.156901}%
  \BibitemOpen
  \bibfield  {author} {\bibinfo {author} {\bibfnamefont {J.}~\bibnamefont {Qian}}, \bibinfo {author} {\bibfnamefont {J.}~\bibnamefont {Li}}, \bibinfo {author} {\bibfnamefont {S.-Y.}\ \bibnamefont {Zhu}}, \bibinfo {author} {\bibfnamefont {J.~Q.}\ \bibnamefont {You}}, \ and\ \bibinfo {author} {\bibfnamefont {Y.-P.}\ \bibnamefont {Wang}},\ }\bibfield  {title} {\enquote {\bibinfo {title} {Probing $pt$-symmetry breaking of non-hermitian topological photonic states via strong photon-magnon coupling}}, }\href {\doibase 10.1103/PhysRevLett.132.156901} {\bibfield  {journal} {\bibinfo  {journal} {Phys. Rev. Lett.}\ }\textbf {\bibinfo {volume} {132}},\ \bibinfo {pages} {156901} (\bibinfo {year} {2024})}\BibitemShut {NoStop}%
\bibitem [{\citenamefont {Feng}\ \emph {et~al.}(2017)\citenamefont {Feng}, \citenamefont {El-Ganainy},\ and\ \citenamefont {Ge}}]{Feng2017}%
  \BibitemOpen
  \bibfield  {author} {\bibinfo {author} {\bibfnamefont {L.}~\bibnamefont {Feng}}, \bibinfo {author} {\bibfnamefont {R.}~\bibnamefont {El-Ganainy}}, \ and\ \bibinfo {author} {\bibfnamefont {L.}~\bibnamefont {Ge}},\ }\bibfield  {title} {\enquote {\bibinfo {title} {Non-hermitian photonics based on parity--time symmetry}}, }\href {\doibase 10.1038/s41566-017-0031-1} {\bibfield  {journal} {\bibinfo  {journal} {Nat. Photonics}\ }\textbf {\bibinfo {volume} {11}},\ \bibinfo {pages} {752} (\bibinfo {year} {2017})}\BibitemShut {NoStop}%
\bibitem [{\citenamefont {Gao}\ \emph {et~al.}(2021)\citenamefont {Gao}, \citenamefont {Xue}, \citenamefont {Gu}, \citenamefont {Liu}, \citenamefont {Zhu},\ and\ \citenamefont {Zhang}}]{Gao2021}%
  \BibitemOpen
  \bibfield  {author} {\bibinfo {author} {\bibfnamefont {H.}~\bibnamefont {Gao}}, \bibinfo {author} {\bibfnamefont {H.}~\bibnamefont {Xue}}, \bibinfo {author} {\bibfnamefont {Z.}~\bibnamefont {Gu}}, \bibinfo {author} {\bibfnamefont {T.}~\bibnamefont {Liu}}, \bibinfo {author} {\bibfnamefont {J.}~\bibnamefont {Zhu}}, \ and\ \bibinfo {author} {\bibfnamefont {B.}~\bibnamefont {Zhang}},\ }\bibfield  {title} {\enquote {\bibinfo {title} {Non-hermitian route to higher-order topology in an acoustic crystal}}, }\href {\doibase 10.1038/s41467-021-22223-y} {\bibfield  {journal} {\bibinfo  {journal} {Nat. Commun.}\ }\textbf {\bibinfo {volume} {12}},\ \bibinfo {pages} {1888} (\bibinfo {year} {2021})}\BibitemShut {NoStop}%
\bibitem [{\citenamefont {Zhang}\ \emph {et~al.}(2021)\citenamefont {Zhang}, \citenamefont {Yang}, \citenamefont {Ge}, \citenamefont {Guan}, \citenamefont {Chen}, \citenamefont {Yan}, \citenamefont {Chen}, \citenamefont {Xi}, \citenamefont {Li}, \citenamefont {Jia}, \citenamefont {Yuan}, \citenamefont {Sun}, \citenamefont {Chen},\ and\ \citenamefont {Zhang}}]{Zhang2021}%
  \BibitemOpen
  \bibfield  {author} {\bibinfo {author} {\bibfnamefont {L.}~\bibnamefont {Zhang}}, \bibinfo {author} {\bibfnamefont {Y.}~\bibnamefont {Yang}}, \bibinfo {author} {\bibfnamefont {Y.}~\bibnamefont {Ge}}, \bibinfo {author} {\bibfnamefont {Y.-J.}\ \bibnamefont {Guan}}, \bibinfo {author} {\bibfnamefont {Q.}~\bibnamefont {Chen}}, \bibinfo {author} {\bibfnamefont {Q.}~\bibnamefont {Yan}}, \bibinfo {author} {\bibfnamefont {F.}~\bibnamefont {Chen}}, \bibinfo {author} {\bibfnamefont {R.}~\bibnamefont {Xi}}, \bibinfo {author} {\bibfnamefont {Y.}~\bibnamefont {Li}}, \bibinfo {author} {\bibfnamefont {D.}~\bibnamefont {Jia}},  \emph {et~al.},\ }\bibfield  {title} {\enquote {\bibinfo {title} {Acoustic non-hermitian skin effect from twisted winding topology}}, }\href {\doibase 10.1038/s41467-021-26619-8} {\bibfield  {journal} {\bibinfo  {journal} {Nat. Commun.}\ }\textbf {\bibinfo {volume} {12}},\ \bibinfo {pages} {6297} (\bibinfo {year} {2021})}\BibitemShut {NoStop}%
\bibitem [{\citenamefont {Wang}\ \emph {et~al.}(2022)\citenamefont {Wang}, \citenamefont {Li}, \citenamefont {Hu}, \citenamefont {Gu}, \citenamefont {Ao}, \citenamefont {Jiang},\ and\ \citenamefont {Gong}}]{PhysRevA.105.023531}%
  \BibitemOpen
  \bibfield  {author} {\bibinfo {author} {\bibfnamefont {X.}~\bibnamefont {Wang}}, \bibinfo {author} {\bibfnamefont {Y.}~\bibnamefont {Li}}, \bibinfo {author} {\bibfnamefont {X.}~\bibnamefont {Hu}}, \bibinfo {author} {\bibfnamefont {R.}~\bibnamefont {Gu}}, \bibinfo {author} {\bibfnamefont {Y.}~\bibnamefont {Ao}}, \bibinfo {author} {\bibfnamefont {P.}~\bibnamefont {Jiang}}, \ and\ \bibinfo {author} {\bibfnamefont {Q.}~\bibnamefont {Gong}},\ }\bibfield  {title} {\enquote {\bibinfo {title} {Non-hermitian high-quality-factor topological photonic crystal cavity}}, }\href {\doibase 10.1103/PhysRevA.105.023531} {\bibfield  {journal} {\bibinfo  {journal} {Phys. Rev. A}\ }\textbf {\bibinfo {volume} {105}},\ \bibinfo {pages} {023531} (\bibinfo {year} {2022})}\BibitemShut {NoStop}%
\bibitem [{\citenamefont {Hu}\ \emph {et~al.}(2023)\citenamefont {Hu}, \citenamefont {Zhang}, \citenamefont {Yue}, \citenamefont {Liao}, \citenamefont {Liu}, \citenamefont {Zhang}, \citenamefont {Cheng}, \citenamefont {Liu},\ and\ \citenamefont {Christensen}}]{PhysRevLett.131.066601}%
  \BibitemOpen
  \bibfield  {author} {\bibinfo {author} {\bibfnamefont {B.}~\bibnamefont {Hu}}, \bibinfo {author} {\bibfnamefont {Z.}~\bibnamefont {Zhang}}, \bibinfo {author} {\bibfnamefont {Z.}~\bibnamefont {Yue}}, \bibinfo {author} {\bibfnamefont {D.}~\bibnamefont {Liao}}, \bibinfo {author} {\bibfnamefont {Y.}~\bibnamefont {Liu}}, \bibinfo {author} {\bibfnamefont {H.}~\bibnamefont {Zhang}}, \bibinfo {author} {\bibfnamefont {Y.}~\bibnamefont {Cheng}}, \bibinfo {author} {\bibfnamefont {X.}~\bibnamefont {Liu}}, \ and\ \bibinfo {author} {\bibfnamefont {J.}~\bibnamefont {Christensen}},\ }\bibfield  {title} {\enquote {\bibinfo {title} {Anti-parity-time symmetry in a su-schrieffer-heeger sonic lattice}}, }\href {\doibase 10.1103/PhysRevLett.131.066601} {\bibfield  {journal} {\bibinfo  {journal} {Phys. Rev. Lett.}\ }\textbf {\bibinfo {volume} {131}},\ \bibinfo {pages} {066601} (\bibinfo {year} {2023})}\BibitemShut {NoStop}%
\end{thebibliography}

\bibliographystyle{apsrev4-1-etal-title}

\end{document}


\title{Supplemental Material for ``Non-Hermitian Quantum Fractals''}
\author{Junsong Sun}
\affiliation{School of Physics, Beihang University,
Beijing, 100191, China}

\author{Chang-An Li}
\email{changan.li@uni-wuerzburg.de}
\affiliation{Institute for Theoretical Physics and Astrophysics, University of W$\ddot{u}$rzburg, 97074 W$\ddot{u}$rzburg, Germany}

\author{Qingyang Guo}
\affiliation{School of Physics, Beihang University,
Beijing, 100191, China}

\author{Weixuan Zhang}
\affiliation{Key Laboratory of advanced optoelectronic quantum architecture and measurements of Ministry of Education, Beijing Key Laboratory of Nanophotonics and Ultrafine Optoelectronic Systems, School of Physics, Beijing Institute of Technology, 100081, Beijing, China}

\author{Shiping Feng}
\affiliation{ Department of Physics,  Beijing Normal University, Beijing, 100875, China}

\author{Xiangdong Zhang}
\email{zhangxd@bit.edu.cn}
\affiliation{Key Laboratory of advanced optoelectronic quantum architecture and measurements of Ministry of Education, Beijing Key Laboratory of Nanophotonics and Ultrafine Optoelectronic Systems, School of Physics, Beijing Institute of Technology, 100081, Beijing, China}

\author{Huaiming Guo}
\email{hmguo@buaa.edu.cn}
\affiliation{School of Physics, Beihang University,
Beijing, 100191, China}

\author{Bj$\ddot{\mathrm{o}}$rn Trauzettel}
\email{trauzettel@physik.uni-wuerzburg.de}
\affiliation{Institute for Theoretical Physics and Astrophysics, University of W$\ddot{u}$rzburg, 97074 W$\ddot{u}$rzburg, Germany}
\date{\today}
\begin{abstract}
In this Supplemental Material, we present and discuss more details
for (S1) Solutions of coupled Hatano-Nelson models on the tree lattice; (S2) Mandelbrot matrix and its characteristic polynomials; (S3) Generalizations of the tree lattice; (S4) Connection of non-Hermitian quantum fractals (NHQFs) to Hofstadter problems with an imaginary magnetic field; (S5) Solutions of NHQFs on a Sierpi$\acute{\mathrm{n}}$ski gasket. 
\end{abstract}
\maketitle
\renewcommand{\thefigure}{S\arabic{figure}}
\renewcommand{\theequation}{S\arabic{equation}}
\renewcommand{\thesection}{S\arabic{section}}
\setcounter{figure}{0}  
\setcounter{equation}{0} 
\numberwithin{equation}{section}\tableofcontents{}

\section{Solutions of coupled Hatano-Nelson models on the tree lattice}
In this section, we present the analytical solution of coupled Hatano-Nelson models on the tree lattice. Let us first present the $\gamma = t$ case in the main text and afterwards present the $\gamma \neq t$ case. 

Specifically, the Hamiltonian for the smallest lattice containing two generations of sites is written as
\begin{equation}\label{eqh2}
  H_2=2t\left(\begin{matrix}
 0&1&0\\
 0&0&1\\
 1&0&0\\
\end{matrix}\right).
\end{equation}
For a system with $n$ generations, the Hamiltonian $H_n$ can be iteratively obtained in terms of $H_{n-1}$:
\begin{equation}
  H_{n}=\left(\begin{matrix}
 0&2t{e_{n-1}}^T&0\\
 0&H_{n-1}&2te_{n-1}{e_{n-1}}^T\\
 2te_{n-1}&0&H_{n-1}\\
\end{matrix}\right), (n\geq3),
\end{equation}
where $e_n$ is a column vector of dimension $2^n-1$ with elements at the first position being 1 and zeros otherwise. $e_n^T$ indicates the transpose of $e_n$.

Let us now search for the eigenenergy of the Hamiltonian. As a starting point, we attempt to solve the eigenenergy from $H_2$. Assuming the eigenstate of $H_2$ is \( |\Psi_2\rangle = \left(\phi_1, \phi_2, \phi_3\right)^T \). According to the eigenvalue equation \( H_2|\Psi_2\rangle=E|\Psi_2\rangle \), we obtain the equations satisfied between components:
\begin{equation}\label{eqh2component}
\begin{split}
\epsilon\phi_1=\phi_2,\\
\epsilon\phi_2=\phi_3,\\
\epsilon\phi_3=\phi_1,
\end{split}
\end{equation}
where $\epsilon=E/2t$. By eliminating the components \( \phi_3\) and \( \phi_2 \), we obtain
\begin{equation}
\begin{split}
\epsilon^2\phi_2&=\phi_1,\\
\epsilon^3\phi_1&=\epsilon^2\phi_2,\\
\epsilon^3\phi_1&=\phi_1.
\end{split}
\end{equation}
Define \( P_2(\epsilon) = \epsilon^3 - 1 \). The eigenenergy of \( H_2 \) is determined by the roots of the characteristic polynomial \( P_2(\epsilon) \). Furthermore, if we treat the wave function component \(\phi_2\) as an undetermined coefficient, the eigenfunction can be represented as \(|\Psi_2\rangle=\phi_2\left(\epsilon^2, 1, \epsilon\right)^T \). 

Next, we seek the eigenenergy spectrum of \( H_3 \). Assuming the eigenstate of \( H_3 \) is 
\begin{equation}
 |\Psi_3\rangle=\left(\phi^3_v,\phi^2_1, \phi^2_2, \phi^2_3,\phi^{2\prime}_1, \phi^{2\prime}_2, \phi^{2\prime}_3 \right)^T, 
\end{equation}
where \( \phi^2_1, \phi^2_2, \phi^2_3 \) are respectively the three wave function components at the three sites of \( H_2 \), and the three primed components represent components at another three sites of \( H_2 \), while \( \phi^3_v \) represents the newly added vertex of \( H_3 \). Therefore, each component of the eigenstate satisfies:
\begin{equation}
\begin{split}
\epsilon\phi^3_v=\phi^2_1,\\
\end{split}
\end{equation}
\begin{equation}\label{eqh21}
\begin{split}
\epsilon\phi^2_1&=\phi^2_2+\phi^{2\prime}_1,\\
\epsilon\phi^2_2&=\phi^2_3,\\
\epsilon\phi^2_3&=\phi^2_1,\\
\end{split}
\end{equation}
\begin{equation}\label{eqh22}
\begin{split}
\epsilon\phi^{2\prime}_1&=\phi^{2\prime}_2+\phi^3_v,\\
\epsilon\phi^{2\prime}_2&=\phi^{2\prime}_3,\\
\epsilon\phi^{2\prime}_3&=\phi^{2\prime}_1.
\end{split}
\end{equation}
We notice that Eq. (\ref{eqh21}) and Eq. (\ref{eqh22}) have the same form as equation Eq. (\ref{eqh2component}) except for the difference in the first line equation. Therefore, we have:
\begin{equation}
\begin{split}
\epsilon^3\phi^2_1&=\epsilon^2\phi^2_2+\epsilon^2\phi^{2\prime}_1\Rightarrow P_2(\epsilon)\phi^2_1=\epsilon^2\phi^{2\prime}_1,\\
\epsilon^2\phi^{2\prime}_1&=\epsilon^2\phi^{2\prime}_2+\epsilon^2\phi^2_1\Rightarrow P_2(\epsilon)\phi^{2\prime}_1=\epsilon^2\phi^3_v,\\
\epsilon\phi^3_v&=\phi^2_1.
\end{split}
\end{equation}
Finally, we can obtain:
\begin{equation}
\begin{split}
P_3(\epsilon)\phi^3_v=[\epsilon P_2(\epsilon)^2-P_1^4(\epsilon)]\phi^3_v=0,
\end{split}
\end{equation}
with $P_1(\epsilon)=\epsilon$. Therefore, we obtain that the eigenenergy spectrum of \( H_3 \) corresponds to all the roots of the characteristic polynomial \( P_3(\epsilon) \). For the form of the eigenfunction \(|\Psi_3\rangle\), we first select \(\phi^2_2\) and \(\phi^{2\prime}_2\) as undetermined coefficients, thus 
\begin{equation}
\begin{split}
|\Psi_3\rangle=\left((\epsilon^3-1)\phi^{2\prime}_2,\epsilon^2\phi^2_2, \phi^2_2, \epsilon\phi^2_2,\epsilon^2\phi^{2\prime}_2, \phi^{2\prime}_2, \epsilon\phi^{2\prime}_2 \right)^T,
\end{split}
\end{equation}
where \(\phi^3_v=(\epsilon^3-1)\phi^{2\prime}_2\) is determined by  Eq. (\ref{eqh22}). Therefore, we also need to know the relationship between \(\phi_2^2\) and \(\phi^{2\prime}_2\) to determine the form of the wave function \(|\Psi_3\rangle\). Through Eq. (\ref{eqh21}) and Eq. (\ref{eqh22}), we can obtain
\begin{equation}
\begin{split}
(\epsilon^3-1)\phi^2_2=\epsilon^2\phi^{2\prime}_2.
\end{split}
\end{equation}
At this point, there is only one undetermined component in \(|\Psi_3\rangle\). Once the eigenvalues are given, the undetermined component can be determined based on the normalization condition.

By repeating the above steps, we find the following iterative relationship among the characteristic polynomials \( P_n(\epsilon) \) of Hamiltonian \( H_n \) of different sizes:
\begin{equation}\label{eqpn}
P_n(\epsilon)=\epsilon P_{n-1}^2(\epsilon)-P_{n-2}^4(\epsilon), \ (n\geq3).
\end{equation}
With this iterative relationship, we can derive the form of any \( P_n(\epsilon) \). Therefore, by solving the characteristic polynomial \( P_n(\epsilon) = 0 \), we can obtain the eigenenergy spectrum of any Hamiltonian \( H_n \). 

The eigenstate of $H_N$ is assumed to be 
\begin{equation}
|\Psi_N\rangle=(\phi^1_1,\phi^{2}_1,\phi^{2}_2,\cdots,\phi^N_1,\cdots,\phi^N_{2^{N-1}})^T,
\end{equation}
where the superscript $i$ and subscript $j$ respectively indicate that this wave function component is located at the $j$-th site of the $i$-th generation. First, we still choose the wave function components $\phi^N_{j\in odd}$ located on odd sites in the $N$-th generation as the undetermined coefficients. Thus, we obtain 
\begin{equation}\label{S115}
\begin{split}
\phi^N_{2j}&=\epsilon\phi^N_{2(j-1)+1}, \\
\phi^{N-1}_j&=\epsilon^2\phi^N_{2(j-1)+1},\\
\phi^{N-2}_j&=\epsilon\phi^{N-1}_{2j}-\phi^{N}_{4(j-1)+3},\\
\phi^{N-3}_j&=\epsilon\phi^{N-2}_{2j}-\phi^{N-1}_{4(j-1)+3},\\
\cdots .
\end{split}
\end{equation}
Once we determine all the wave function components \(\phi^N_{2(j-1)+1}\) on the odd lattice sites of the $N$-th generation, we can then derive the form of all the remaining wave function components according to the above rule [Eq. (\ref{S115})]. Since there are \(2^{N-2}\) undetermined \(\phi^N_{2(j-1)+1}\), we need to express them as undetermined coefficients based on the relationships between these undetermined components. Then, according to the normalization condition of the wave function, we can determine all the wave function components. And these undetermined wave function components have the following relationship:
\begin{equation}\label{eqphin}
\begin{split}
P_2(\epsilon)\phi^N_{4(j-1)+1}&=P_1^2(\epsilon)\phi^N_{4(j-1)+3}, \\
P_3(\epsilon)\phi^N_{8(j-1)+3}&=P_2^2(\epsilon)\phi^N_{8(j-1)+7}, \\
P_4(\epsilon)\phi^N_{16(j-1)+7}&=P_3^2(\epsilon)\phi^N_{16(j-1)+15},\\
\cdots\\
P_{N-1}(\epsilon)\phi^N_{2^{N-1}(j-1)+2^{N-2}-1}&=P_{N-2}^2(\epsilon)\phi^N_{2^{N-1}(j-1)+2^{N-1}-1}.
\end{split}
\end{equation}
Therefore, we can express all the wave function components using one undetermined component, thus, obtaining the form of the entire wave function.

Now, we present the analytical solution of the eigenvalue problem for the general condition when \(\gamma \neq t\).  
We first seek the solution to the Hamiltonian $H_2$ in Eq.~(2) of the main text. Assuming the eigenfunction of $H_2$ is $|\Psi_2\rangle = \left[\phi_1, \phi_2, \phi_3\right]^T$ and expanding the eigen-equation $H_2|\Psi_2\rangle=E|\Psi_2\rangle$, we obtain the following equations satisfied by the components:
\begin{equation}\label{aeqh2component}
\begin{split}
E\phi_1=(t+\gamma)\phi_2+(t-\gamma)\phi_3,\\
E\phi_2=(t+\gamma)\phi_3+(t-\gamma)\phi_1,\\
E\phi_3=(t+\gamma)\phi_1+(t-\gamma)\phi_2.
\end{split}
\end{equation}
By eliminating the components \( \phi_3\) and \( \phi_2 \), we obtain
\begin{equation}
\begin{split}
\alpha_2\phi_2&=\beta_2^-\phi_1,\\
\alpha_2\phi_3&=\beta_2^+\phi_1,\\
E\alpha_2\phi_1&=\left[(t+\gamma)^3+(t-\gamma)^3+2E(t^2-\gamma^2)\right]\phi_1,
\end{split}
\end{equation}
where 
\begin{equation}
\begin{split}
\alpha_2&=E^2-(t^2-\gamma^2),\\
\beta_2^-&=\left[(t+\gamma)^2+E(t-\gamma)\right],\\
\beta_2^+&=\left[(t-\gamma)^2+E(t+\gamma)\right].
\end{split}
\end{equation}
We define the characteristic polynomial
\begin{equation}
\begin{split}
P_2 &= E\alpha_2-\left[(t+\gamma)^3+(t-\gamma)^3+2E(t^2-\gamma^2)\right],
\end{split}
\end{equation}
and the eigenvalues of $H_2$ are the roots of $P_2=0$. 
Additionally, the eigenfunction can be expressed in terms of $\phi_1$ as  
\begin{equation}
\begin{split}
|\Psi_2\rangle=\phi_1\left(1,\frac{\beta_2^-}{\alpha_2},\frac{\beta_2^+}{\alpha_2}\right)^T.
\end{split}
\end{equation}
Next, we proceed to solve the eigenvalue problem for $H_3$. The eigenfunction is assumed to be 
\begin{equation}
 |\Psi_3\rangle=\left(\phi^3_v,\phi^2_1, \phi^2_2, \phi^2_3,\phi^{2\prime}_1, \phi^{2\prime}_2, \phi^{2\prime}_3 \right)^T,
\end{equation}
where \( \phi^2_1, \phi^2_2, \phi^2_3 \) (and \( \phi^{2\prime}_1, \phi^{2\prime}_2, \phi^{2\prime}_3 \)) are the three components on the sites forming the left~(right)-bottom triangle, and \( \phi^3_v \) represents the component on the uppermost site. Expanding the eigen-equation $H_3|\Psi_3\rangle=E|\Psi_3\rangle$, we obtain the following set of equations:
\begin{equation}
\begin{split}
E\phi^3_v=(t+\gamma)\phi^2_1+(t-\gamma)\phi^{2\prime}_1,\\
\end{split}
\end{equation}
\begin{equation}\label{aeqh21}
\begin{split}
E\phi^2_1&=(t+\gamma)\phi^2_2+(t-\gamma)\phi^2_3
+(t+\gamma)\phi^{2\prime}_1+(t-\gamma)\phi^3_v,\\
E\phi^2_2&=(t+\gamma)\phi^2_3+(t-\gamma)\phi^2_1,\\
E\phi^2_3&=(t+\gamma)\phi^2_1+(t-\gamma)\phi^2_2,\\
\end{split}
\end{equation}
\begin{equation}\label{aeqh22}
\begin{split}
E\phi^{2\prime}_1&=(t+\gamma)\phi^{2\prime}_2+(t-\gamma)\phi^{2\prime}_3
+(t+\gamma)\phi^3_v+(t-\gamma)\phi^2_1,\\
E\phi^{2\prime}_2&=(t+\gamma)\phi^{2\prime}_3+(t-\gamma)\phi^{2\prime}_1,\\
E\phi^{2\prime}_3&=(t+\gamma)\phi^{2\prime}_1+(t-\gamma)\phi^{2\prime}_2.
\end{split}
\end{equation}
It is noticed that the equations have the same form except for the first ones in Eq. (\ref{aeqh21}) and Eq. (\ref{aeqh22}). By appropriately eliminating the other components, the set of linear equations for \(\phi^2_1, \phi^{2\prime}_1, \phi^3_v\) can be written as:
\begin{equation}
\begin{split}
P_2\phi^2_1&=\alpha_2\left[(t+\gamma)\phi^{2\prime}_1+(t-\gamma)\phi^3_v\right],\\
P_2\phi^{2\prime}_1&=\alpha_2\left[(t+\gamma)\phi^3_v+(t-\gamma)\phi^2_1\right],\\
E\phi^3_v&=(t+\gamma)\phi^2_1+(t-\gamma)\phi^{2\prime}_1.\\
\end{split}
\end{equation}
Moreover, the right-hand sides of the above equations can all be expressed in terms of \(\phi^3_v\):
\begin{equation}
\begin{split}
\alpha_3\phi^2_1&=\beta_3^-\phi^3_v,\\
\alpha_3\phi^{2\prime}_1&=\beta_3^+\phi^3_v,\\
E\alpha_3\phi^3_v&=\left\{\alpha_2^2\left[(t+\gamma)^3+(t-\gamma)^3\right]+2\alpha_2P_2(t^2-\gamma^2)\right\}\phi^3_v,\\
\end{split}
\end{equation}
where 
\begin{equation}
\begin{split}
\alpha_3&=p_2^2-\alpha_2^2(t^2-\gamma^2),\\
\beta_3^-&=\left[\alpha_2p_2(t-\gamma)+\alpha_2^2(t+\gamma)^2\right],\\
\beta_3^+&=\left[\alpha_2p_2(t+\gamma)+\alpha_2^2(t-\gamma)^2\right].
\end{split}
\end{equation}
Similarly, we define the characteristic polynomial of $H_3$ as:
\begin{equation}
\begin{split}
P_3=E\alpha_3-\left\{\alpha_2^2\left[(t+\gamma)^3+(t-\gamma)^3\right]+2\alpha_2P_2(t^2-\gamma^2)\right\}\nonumber.
\end{split}
\end{equation}
The eigenvalues of $H_3$ can be obtained by solving the cubic equation $P_3=0$. 
In terms of $\phi^3_v$, the eigenfunction can be expressed as
\begin{equation}
\begin{split}
|\Psi_3\rangle=\phi^3_v\left(1, \frac{\beta_3^-}{\alpha_3},\frac{\beta_3^-\beta_2^-}{\alpha_3\alpha_2},\frac{\beta_3^-\beta_2^+}{\alpha_3\alpha_2},\frac{\beta_3^+}{\alpha_3},\frac{\beta_3^+\beta_2^-}{\alpha_3\alpha_2},\frac{\beta_3^+\beta_2^+}{\alpha_3\alpha_2}\right)^T.
\end{split}
\end{equation}

\begin{figure}[hbpt]
  \includegraphics[width=8.7cm]{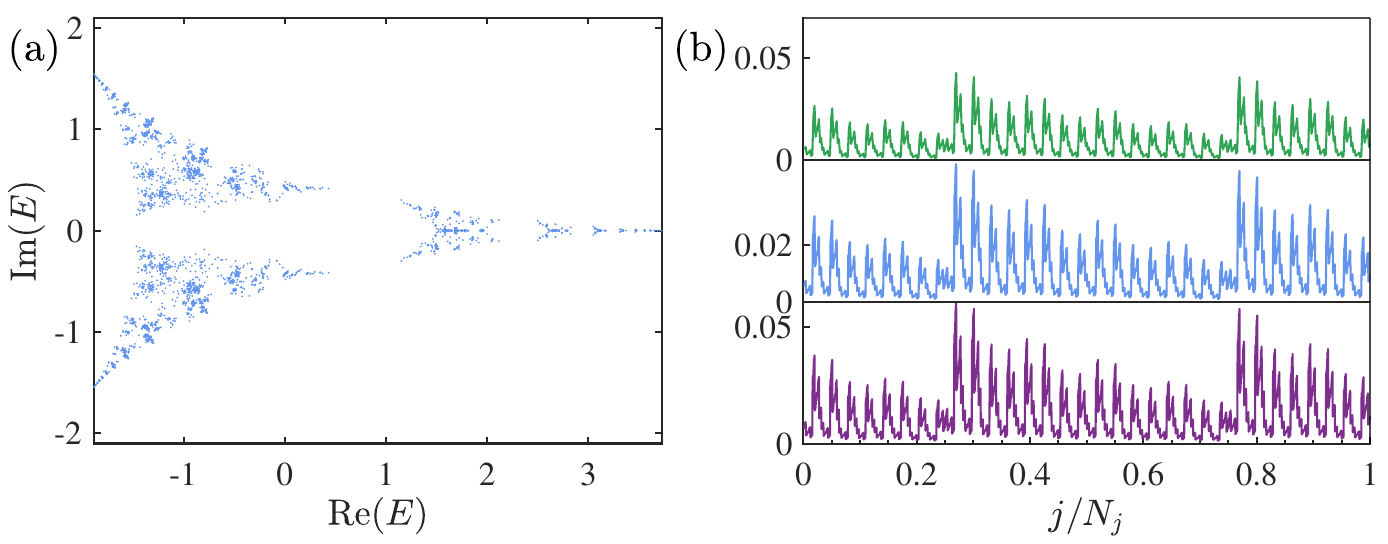}
  \caption{(a) The complex eigenvalues of Hatano-Nelson model when $\gamma \neq t$ on a binary tree lattice comprising $N=11$ generations. (b) The amplitude of the wave function in the three lowest generations as a function of the scaled index $j/N_i$. Here the parameter $\gamma/t=0.5$ is used.}\label{afig1}
\end{figure}

By repeating the above procedure, we find the following iterative relationship for the characteristic polynomials $P_n$ of the Hamiltonian $H_n$:
\begin{equation}\label{aeqpn}
\begin{split}
P_n=&E\alpha_n-\left\{\alpha_{n-1}^2\left[(t+\gamma)^3+(t-\gamma)^3\right]
+2\alpha_{n-1}P_{n-1}(t^2-\gamma^2)\right\}, 
\end{split}
\end{equation}
where $\alpha_n=P_{n-1}^2-\alpha_{n-1}^2(t^2-\gamma^2)$. By solving the equation $P_n = 0$, we can obtain the energy spectrum of the Hamiltonian $H_n$. 
Generally, the eigenstate of $H_n$ can be written as,
\begin{equation}
|\Psi_n\rangle=\left(\phi^1_1,\phi^{2}_1,\phi^{2}_2,\cdots,\phi^n_1,\cdots,\phi^n_{2^{n-1}}\right)^T,
\end{equation}
where the superscript $i$ and subscript $j$ denote the generation index and site index within each generation, respectively. The relations among the components are as follows: 
\begin{equation}\label{aeqphi}
\begin{split}
\phi^1_{1}&=\phi^1_{1}, \\
\phi^{2}_{j}&=\frac{\beta_n^{\pm}}{\alpha_n}\phi_1^1,\\
\phi^{3}_j&=\frac{\beta^{\pm}_n\beta^\pm_{n-1}}{\alpha_n \alpha_{n-1}}\phi^1_1,\\
&\cdots, \\
\phi^{n}_j&=\frac{\beta^{\pm}_n\beta^\pm_{n-1} \cdots \beta_1^{\pm}}{\alpha_n \alpha_{n-1}\cdots \alpha_1}\phi^1_1,\\
\end{split}
\end{equation}
where 
\begin{equation}\label{aeqphi}
\begin{split}
\beta_n^-=\alpha_{n-1}P_{n-1}(t-\gamma)+\alpha_{n-1}^2(t+\gamma)^2,\\
\beta_n^+=\alpha_{n-1}P_{n-1}(t+\gamma)+\alpha_{n-1}^2(t-\gamma)^2.
\end{split}
\end{equation}
Therefore, the wave function can be expressed in terms of $\phi^1_{1}$. Finally, it is uniquely determined by the normalization condition.

Figure \ref{afig1} illustrates the energy spectrum at $\gamma/t=0.5$. It is evident that the pattern is stretched along the $\mathrm{Re}(E)$ direction, and the three-fold rotational symmetry is broken. The pseudo-Hermitian symmetry $\eta\mathcal{H}\eta=\mathcal{H}^\dagger$ suggests that both $E$ and its conjugate $E^{*}$ are eigenvalues. Consequently, the energy spectrum exhibits a mirror symmetry about the $\mathrm{Im}(E)$=0. The fractal characteristics is still visible in the energy spectrum, particularly pronounced in the horizontal branch. Furthermore, the wave function displays self-similarity, similar to the case of $\gamma=t$.


\section{Mandelbrot matrix and its characteristic polynomials}
In this section, we review the Mandelbrot matrix and its characteristic polynomials following Ref. \cite{afraceigenvector}. We show that the Hamiltonian $H_n$ describing the coupled Hatano-Nelson model on a tree lattice satisfies the basic facts of the Mandelbrot matrix. 

The Mandelbrot matrix $M_n$ is defined in an iterative way. Explicitly, the starting point is
\begin{equation}
    M_1=(1). 
\end{equation}
Then, one puts 
\begin{equation}\label{}
  M_{2}=\left(\begin{matrix}
 M_1&0&1\\
 1&0&0\\
 0&1&M_1\\
\end{matrix}\right).
\end{equation}
Proceeding in a similar way, one can construct $M_{n+1}$ from $M_n$ as 
\begin{equation}\label{}
  M_{n+1}=\left(\begin{matrix}
 M_{n}&\mathbf{0}&e_{n}e_{d_n}^T\\
 e_{d_n}^T&\mathbf{0}&\mathbf{0}\\
 \mathbf{0}&e_n&M_{n}\\
\end{matrix}\right),
\end{equation}
where $e_{n}=(1,0,0,\cdots,0)^T$ is the leading elementary column vector of dimension $2^n-1$ and $e_{d_n}$ is the final elementary column vector of the same dimension. 

The Mandelbrot matrix has some fundamental properties as follows:
\begin{itemize}
    \item The matrix $M_n$ has dimension $2^n-1$;
    \item The matrix $M_n$ has determinant $\mathrm{det}M_n$=1 for all $n\geq 1$;
    \item The matrix $M_n$ are all upper Hessenberg: that is they are upper triangular except that principal sub-diagonal is also nonzero with zeros and ones.
\end{itemize}
These basic facts of Mandelbrot matrix also fit the Hamiltonian matrix $H_n$ proposed for describing coupled Hatano-Nelson modelds in the main text. 

The defined Mandelbrot matrix satisfies the relation $\mathrm{det}(\lambda \mathbf{I}+ M_n)=P_{n+1}(\lambda)$ where $P_{n}(\lambda)$ is the characteristic polynomials of Mandelbrot matrix. It satisfies the recurrence relation 
\begin{equation}
    P_{n+1}(\lambda)=\lambda P_{n}^2(\lambda)+1.
\end{equation}
This recurrence relation is actually a variation of the Mandelbrot fundamental recurrence $z_{n+1}=z_n^2+c$. One can identify these two by putting $P_n=z_n/c$ and relabel $c$ to be $\lambda$. Zeros of the characteristic polynomials give periodic points in the Mandelbrot set. 

For our case, Eq. (4) in the main text represents a variation of the recurrence of Mandelbrot polynomials presented above. Thus the NHQFs obtained from  Eq. (4) in the main text follow the Mandelbrot-like set in fractal theory.

\section{Generalizations of the tree lattice}
In this section, we generalize the tree lattice  to accommodate an arbitrary number ($q$) of branches. 
 The iterative relation of the characteristic polynomials for determining the eigenvalues at the case $\gamma=t$ is:
\begin{equation}\label{eqpnzq}
\begin{split}
P_n(\epsilon)=\epsilon P_{n-1}^q(\epsilon)-P_{n-2}^{q^2}(\epsilon),~(n\geq3),
\end{split}
\end{equation}
where $P_2(\epsilon)=\epsilon^{q+1}-1$.
The $q+1$ power of $\epsilon$ in $P_2(\epsilon)$ restricts the energy spectrum to exhibit $(q+1)$-fold rotational symmetry about the origin. 
Assuming the form of the wave function to be $|\Psi_n\rangle=\left(\phi^1_1,\phi^{2}_1,\cdots,\phi^{2}_q,\cdots,\phi^n_1,\cdots,\phi^n_{q^{n-1}}\right)^{T}$,  
we can derive the following recurrence relation for the components of the wave function: 
\begin{subequations}\label{eqphi}
\begin{align}
&\phi^n_{q(j-1)+1+t}=\epsilon^t\phi^n_{q(j-1)+1}~(1\leq t <q),\\
&\phi^{n-1}_j=\epsilon^q\phi^n_{q(j-1)+1},\\
&\phi^{n-k}_j=\epsilon\phi^{n-k+1}_{qj}-\phi^{n-k+2}_{q^2(j-1)+q^2-q+1}~(1<k<n),
\end{align}
\end{subequations}
which, together with the normalization condition, allows us to determine the wave functions.

\begin{figure}[hbpt]
  \includegraphics[width=8.6cm]{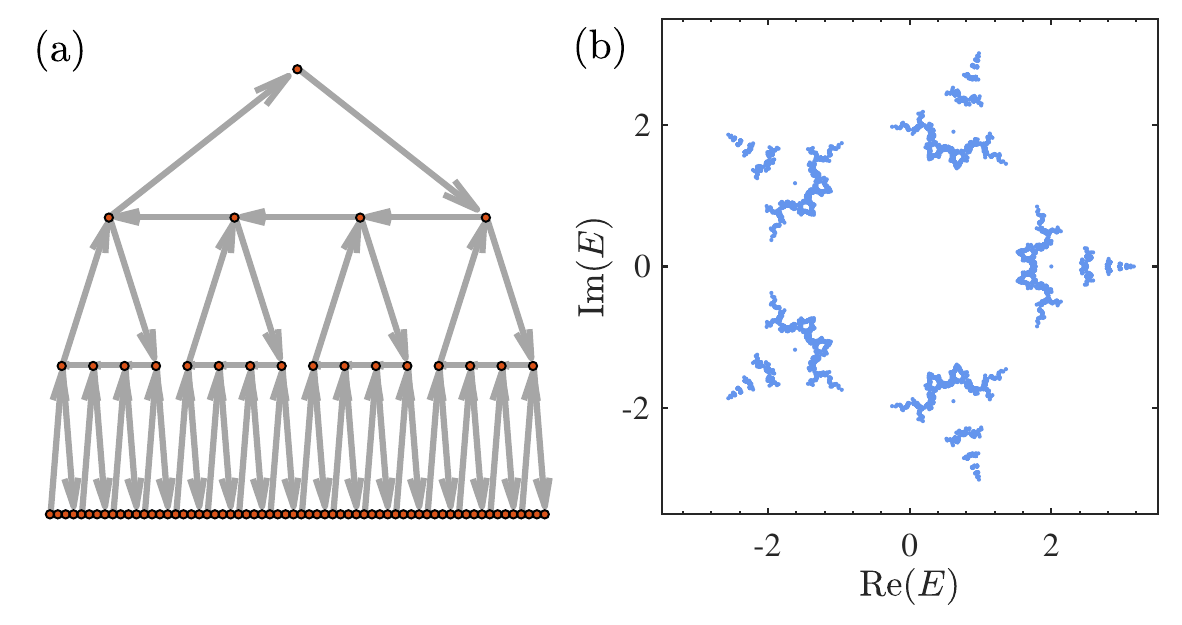}
  \caption{(a) An extension of the binary tree lattice to accommodate four branches. (b) The eigenvalues of Hatano-Nelson model on the extended tree lattice in (a) with $N=7$ generations. The pattern in (b) displays a five-fold rotation
symmetry about the origin.}\label{afig2}
\end{figure}

\begin{figure}[hbpt]
  \includegraphics[width=8.6cm]{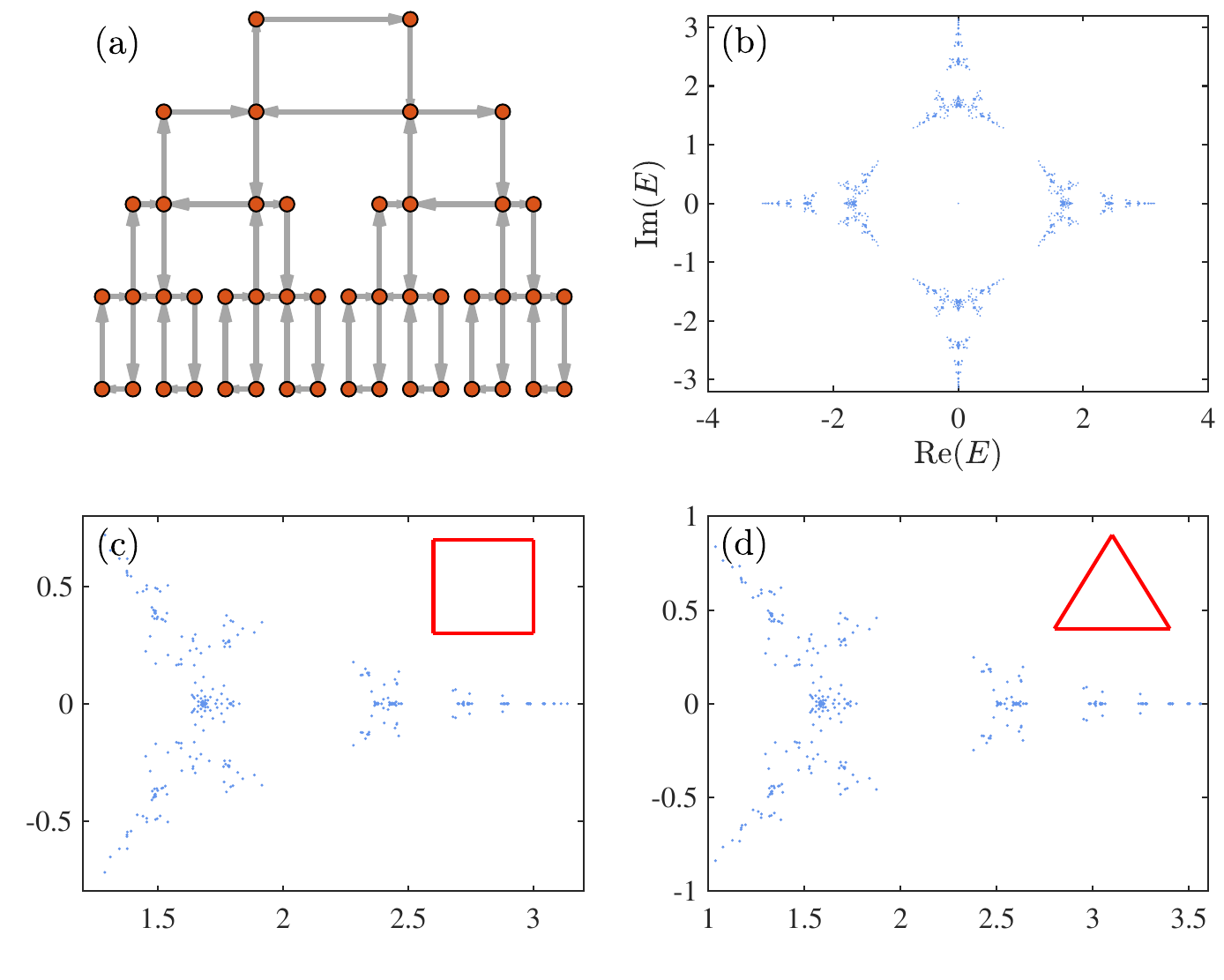}
  \caption{(a) An alternative extension of the binary tree lattice involving replacing the original triangular structures with polygons. (b) The eigenvalues of Hatano-Nelson model on the geometry in (a), whose pattern exhibits a fourfold rotational symmetry. (c) The enlarged plot of one branch in (b). For comparison purpose, (d) shows the enlarged plot of one branch in Fig.~1(b) of the main text. In (b), (c) and (d), the energy spectra are calculated on geometries with $N=10$ generations. }\label{afig3}
\end{figure}

In the main text, we have chosen $q = 3$ as an example, and shown the energy spectrum in the complex plane. To further observe the evolution of the energy-spectrum pattern with the branch number $q$, we compute the energy spectrum on the fractal-tree lattice with $q=4$ branches. As expected, the energy spectrum displays five-fold rotational symmetry, confirming the statement that the energy spectrum of the fractal-tree lattice with $q$ branches will exhibit ($q + 1$)-fold rotational symmetry about the origin.

There are also alternative methods for extending the binary tree structure to generate an energy spectrum with multi-fold rotational symmetry. One such method involves replacing the original triangular structures with polygons [such as quadrilaterals shown in Fig.~\ref{afig3}]. In this manner, the energy spectrum of the resulting fractal structure demonstrates fourfold rotational symmetry around the origin, while maintaining the self-similar structure within each branch. The similar results lie in the fact that the Hamiltonians and the associated characteristic polynomials here also satisfy recursive relations.
The characteristic polynomial of the Hamiltonian for the simplest case, containing one quadrilateral (four sites), is $P_2(\epsilon) = \epsilon^4 - 1$. Furthermore, the characteristic polynomial of the Hamiltonian for the geometry with three generations is $P_3(\epsilon) = \epsilon^2(P_2^2(\epsilon) - \epsilon^4)$. By further derivation, the characteristic polynomial $P_n$ of the Hamiltonian $H_n$ has the following iterative relation: 
\begin{equation}\label{eqb1}
\begin{split}
P_n(\epsilon) = P_1^2(\epsilon)[P_{n-1}^2(\epsilon) - P_{n-2}^4(\epsilon)],
\end{split}
\end{equation}
where $P_1(\epsilon)=\epsilon$. It is found that the fractal pattern here has a correspondence to that shown in Fig.~1(b) of the main text for a separate branch, which can be understood by analyzing the iterative relations given in Eq.~(\ref{eqb1}) and Eq.~(4). Specifically, if we treat both $\epsilon^3$ in Eq.~(4) and $\epsilon^4$ in Eq.~(\ref{eqb1}) as the same new variable \( \varepsilon \), we can observe that Eq. (\ref{eqb1}) and Eq. (4) become identical except for a global coefficient, thus sharing the same roots for $\varepsilon$. Therefore, for geometries with the same number of generations, each eigenvalue $E$ in Fig.~\ref{afig3}(c) corresponds to one $E'$ in Fig.~\ref{afig3}(d), related by the equation $(E/2t)^4=(E'/2t)^3$.

\section{Connection of non-Hermitian quantum fractals to Hofstadter problems with an imaginary magnetic field}
\begin{figure}[hbpt]
  \includegraphics[width=8.6cm]{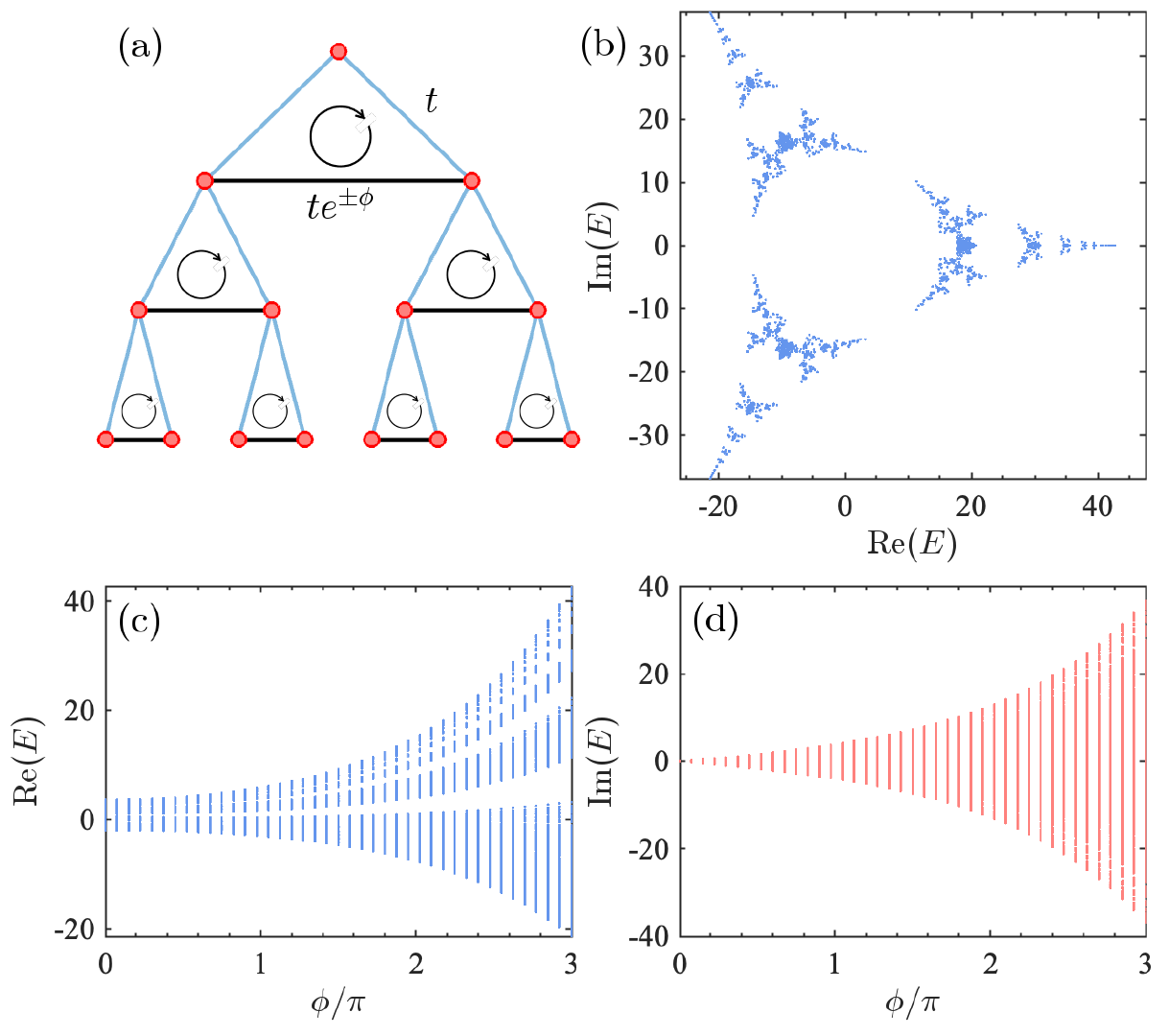}
  \caption{(a) Schematic diagram of a tight-binding model on a tree lattice under an imaginary magnetic field. The hopping amplitudes along the black bonds carry an imaginary phase $e^{\pm \phi}$, where $\phi$ is positive (negative) when hopping to the left (right). (b) Energy spectrum under an imaginary magnetic field with an imaginary flux $\phi = 3\pi$ for a lattice spanning $N = 12$ generations. (c) and (d) represent the real and imaginary parts of the spectrum as functions of the flux of the imaginary magnetic field, respectively.}\label{afigimf}
\end{figure}

In this section, we show the connection of NHQFs to Hofstadter problems with an imaginary magnetic field. A magnetic field is incorporated into a tight-binding model using Peierls substitution in the hopping amplitudes. The corresponding Hamiltonian under an imaginary magnetic field can be written as
\begin{equation}
  \mathcal{H}_\phi=\sum_{{\langle ij\rangle}}\left[t_{ij}c_i^\dagger c_j+t_{ij}^\dagger c_j^\dagger c_i\right],
\end{equation}
where the hopping amplitude in the horizontal black bond is $t_{ij} = te^{{\rm i} (\pm{\rm i}\phi)}$, and \( t_{ij} = t \) for other bonds [see Fig. \ref{afigimf}(a)]. This gauge choice ensures that an electron hopping clockwise around each triangular loop acquires a phase of \( {\rm +i}\phi \), corresponding to the flux of the imaginary magnetic field. 

It is observed that as the strength of the imaginary magnetic field increases, the spectrum of the system closely resembles that of the Hatano-Nelson model  at \(  \gamma=t \), as described in the main text. As shown in Fig. \ref{afigimf}(b), when the phase is large (such as \( \phi = 3\pi \)), the pattern of the spectrum is nearly identical to that in the main text. The NHQFs in energy spectrum follow the Mandelbrot-like set in fractal theory. For Hofstadter problems at a single flux value, the energy spectrum depends on the specific value of flux. For rational flux value $\phi= p/q$, the spectrum consists of $q$ sub-bands. Whereas for irrational flux values, the spectrum is an infinite Cantor set.

We also illustrate the real and imaginary parts of all eigenvalues as a function of flux \( \phi \) in Figs. \ref{afigimf}(c) and (d), respectively, and observe no signature of Hofstadter butterfly in this non-Hermitian cases.

\section{Solutions of non-Hermitian quantum fractals on a Sierpi$\acute{\mathrm{n}}$ski gasket}
In this section, we present the details of the analytical solution of Hatano-Nelson model on a Sierpi$\acute{\mathrm{n}}$ski gasket. For the smallest lattice containing three sites, the Hamiltonian can be expressed as
\begin{equation}\label{}
  H_1=2t\left(\begin{matrix}
 0&1&0\\
 0&0&1\\
 1&0&0\\
\end{matrix}\right).
\end{equation}
Then for a general Sierpi$\acute{\mathrm{n}}$ski gasket with $n$ generations, the Hamiltonian $H_n$ can be written in terms of $H_{n-1}$,
\begin{equation}\label{}
  H_{n}=\left(\begin{matrix}
 H_{n-1}&2tC&0\\
 0&H_{n-1}&2tC\\
 2tC&0&H_{n-1}\\
\end{matrix}\right),
\end{equation}
where $C$ denotes the connection matrix between $H_{n-1}$. The matrix $C$ contains only one non-zero element, whose value is $1$, located at the $(D_{n-1}/3 + 1)$-th row and $(2D_{n-1}/3 + 1)$-th column (where $D_{n-1}$ is the matrix dimension of $H_{n-1}$).

The energy spectrum and wave function can be directly obtained from the eigenvalue problem of $H_1$:
\begin{equation}
\begin{split}
    \epsilon\phi_1&=\phi_2,\\
    \epsilon\phi_2&=\phi_3,\\
    \epsilon\phi_3&=\phi_1,
\end{split}
\end{equation}
with $\epsilon=E/2t$. Therefore, the eigenvalue $\epsilon$ of $H_1$ satisfies $\epsilon^3-1=0$ and its corresponding wave function is $|\Psi_1\rangle=\phi_1(1, \epsilon, 1/\epsilon) $. We define $P_1(\epsilon) = \epsilon^3 - 1$, which represents the characteristic polynomial of $H_1$. The roots of $P_1(\epsilon) = 0$ determine the energy spectrum.

For the case of $n=2$, the eigenvalue problem $H_2|\Psi_2\rangle=E|\Psi_2\rangle$ leads to the following set of equations for the components:
\begin{equation}
\begin{split}
    \phi^x_1&=\phi^x_1,\\
    \epsilon\phi^x_1&=\phi^x_2,\\
    \epsilon\phi^x_3&=\phi^x_1,\\
\end{split}
\end{equation}
where $x=a, b, c$ respectively denote the three sub-triangular regions into which the entire lattice is divided. It is evident that the threefold rotational symmetry of the system ensures that the wave function takes the same form in each of the three sub-triangles when parameterized with the top vertex of each sub-triangle, and the magnitudes at the top vertices of the three sub-triangles must be equal, i.e., $|\phi^a_1|=|\phi^b_1|=|\phi^c_1|$. Therefore, the magnitudes of the wave functions in the three sub-triangles must be identical. 

\begin{figure*}[hbpt]
  \includegraphics[width=16.8cm]{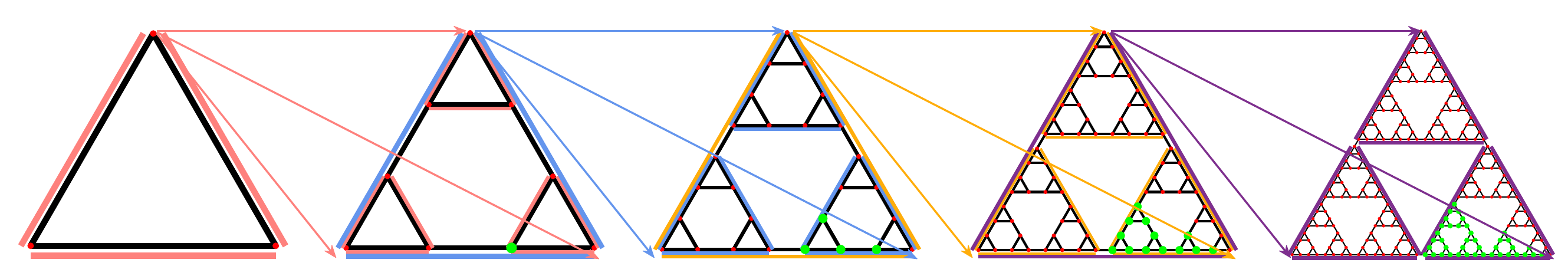}
  \caption{A schematic illustration of the self-similarity of the wave functions of the Hatano-Nelson model on the Sierpi$\acute{\mathrm{n}}$ski gasket. The magnitudes of the wave function within triangles of the same color differing only by an overall scale factor. Due to the threefold rotational symmetry of the system, we show only the top sub-triangle for each lattice size here. The number of generations are $N=2,3,4,5,6$ from left to right, respectively. The green dots indicate sites where the wave function is not simply proportional to those in earlier generations.}\label{afig5}
\end{figure*}

The top vertices of the three sub-triangles hold equivalent positions, and their values are linked by the characteristic polynomial of $H_1$:
\begin{equation}
\begin{split}
    P_1(\epsilon)\phi^a_1=\phi^b_1,\\
    P_1(\epsilon)\phi^b_1=\phi^c_1,\\
    P_1(\epsilon)\phi^c_1=\phi^a_1.
\end{split}
\end{equation}
Therefore, we derive the characteristic polynomial of $H_2$, given by $P_2(\epsilon) = P_1^3(\epsilon) - 1$. And the wave function of $H_2$ can be expressed as 
\begin{equation}
\begin{split}
|\Psi_2\rangle=\left(|\Psi_1\rangle, P_1| \Psi_1\rangle, \frac{1}{P_1}|\Psi_1\rangle\right).
\end{split}
\end{equation}

Similarly, for a general lattice with $n$ generations, we can divide the entire lattice into three sub-triangles labeled as $a,b$ and $c$, and the wave function of $H_n$ can be written as
\begin{equation}
\begin{split}
|\Psi_n\rangle=\left(|\Psi^a_n\rangle, \frac{P_{n-1}}{\alpha_{n-1}}|\Psi^a_n\rangle, \frac{\alpha_{n-1}}{P_{n-1}}|\Psi^a_n\rangle\right),
\end{split}
\end{equation}
where $\alpha^3_{n-1}=\prod_{i=0}^{n-3} P_i^6(\epsilon)\left[P_{n-2}^3(\epsilon)-P_{n-1}(\epsilon)\right]^2$. 
The wave function $|\Psi^a_n\rangle$ in the sub-triangle $a$ is recursively determined by the wave function $|\Psi^a_{n-1}\rangle$ in its corresponding sub-triangle $a$ from the previous generation:
\begin{equation}\label{eqPsina}
\begin{split}
|\Psi^a_n\rangle=\left(|\Psi^a_{n-1}\rangle, (\frac{P_{n-2}}{\alpha_{n-2}}|\Psi^a_{n-1}\rangle)_{\rm part}, \frac{\alpha_{n-2}}{P_{n-2}}|\Psi^a_{n-1}\rangle\right)
\end{split}
\end{equation}
In Eq.~(\ref{eqPsina}), the subscript 'part' in the second term on the right-hand side indicates that in the second sub-triangle, only the components on part of the lattice sites follow the proportional relation. There are lattice sites that do not follow this relation, marked by green dots in Fig.~\ref{afig5}. 
\begin{figure}[hbpt]
  \includegraphics[width=8.6cm]{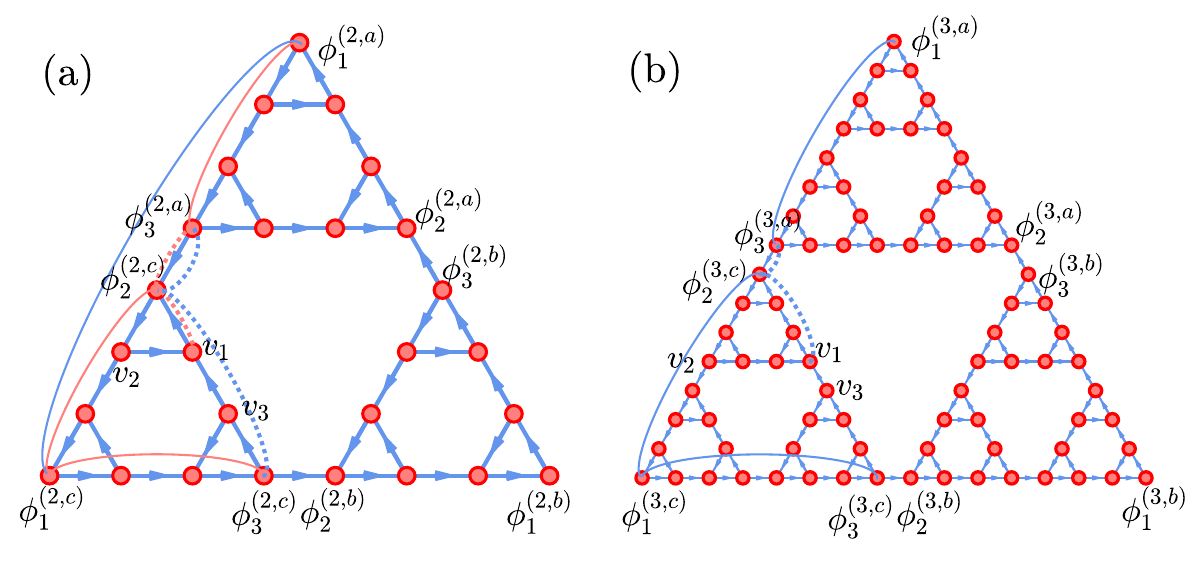}
  \caption{Schematic demonstration of the derivations for (a) Eq.~(\ref{c3a}) with $N=3$ generations and (b) Eq. (\ref{c15a}) with $N=4$ generations.  }\label{afig6}
\end{figure}

To establish the recursive formula of the wave function in Eq.~(\ref{eqPsina}), it is crucial to know the relationships among the components at the three vertices of the triangular system. These relationships will also determine the characteristic polynomial of $H_n$. We denote the three components of each sub-triangle as $\phi^{(n-1,x)}_i$, where $n$ represents the number of generations of the lattice; $x=a,b,c$ denotes the three sub-triangular regions in the lattice with $n$ generations; $i=1,2,3$ represents the three vertices of each sub-triangle with $n-1$ generations. The sequence $x = a, b, c$ and $i = 1, 2, 3$ follows a clockwise order, which is opposite to the direction of the nonreciprocal hoppings. Figure \ref{afig5} displays the $n=2$ Sierpi$\acute{\mathrm{n}}$ski gasket, and the related nine components can be written as 
\begin{equation}
  \phi^{1,x}_1,\ \phi^{1,x}_2,\ \phi^{1,x}_3\ \ (x=a,b,c).
\end{equation}
From the eigenvalue problem $H_2|\Psi_2\rangle=E|\Psi_2\rangle$, we can obtain the following set of equations:
\begin{equation}
\begin{split}
 \epsilon\phi^{(1,c)}_1&=\phi^{(1,c)}_2,\\
 \epsilon\phi^{(1,c)}_2&=\phi^{(1,a)}_3+\phi^{(1,c)}_3,\\
 \epsilon\phi^{(1,a)}_3&=\phi^{(1,a)}_1,\\
 \epsilon\phi^{(1,c)}_3&=\phi^{(1,c)}_1.
\end{split}
\end{equation}
Then we have
\begin{equation}
\begin{split}
&\epsilon^2\phi^{(1,c)}_1=\epsilon\phi^{(1,c)}_2=\phi^{(1,a)}_3+\phi^{(1,c)}_3,\\
&\epsilon^3\phi^{(1,c)}_1=\epsilon\phi^{(1,a)}_3+\epsilon\phi^{(1,c)}_3=\phi^{(1,c)}_1+\phi^{(1,a)}_1,\\
&(\epsilon^3-1)\phi^{(1,c)}_1=P_1(\epsilon)\phi^{(1,c)}_1=\phi^{(1,a)}_1.
\end{split}
\end{equation}
Similarly, we can obtained $P_1(\epsilon)\phi^{(1,a)}_1=\phi^{(1,b)}_1$ and $P_1(\epsilon)\phi^{(1,b)}_1=\phi^{(1,c)}_1$, which reflect the threefold rotation symmetry of the system. We then derive the characteristic polynomial of $H_2$, given by $P_2(\epsilon) = P_1^3(\epsilon) - 1$.

We notice that this relationship is still satisfied in next generation of lattice except $P_1(\epsilon)\phi^{(1,b)}_1=\phi^{(1,c)}_1$ broken. This is because on the lattice of generation $n=3$, the site corresponding to \(\phi^{(1,b)}_1\) has an additional in-arrow, which will establish connections within the wavefunction components at another $n=2$ generation triangles. However, the additional out-arrow on \(\phi^{(1,c)}_1\) will not have this effect.

Furthermore, for the case of $n=3$ [see Fig. \ref{afig6}(a)], we have
\begin{equation}
\begin{split}
P_1(\epsilon)\phi^{(2,c)}_1&=\phi^{(2,c)}_2,\\
P_1(\epsilon)\phi^{(2,c)}_3&=\phi^{(2,c)}_1,\\
P_1(\epsilon)\phi^{(2,a)}_3&=\phi^{(2,a)}_1,\\
P_1(\epsilon)\phi^{(2,c)}_2&=\phi^{(2,c)}_3+\epsilon^2\phi^{(2,a)}_3.
\end{split}
\end{equation}
The derivation of the final equation above is as follows:
\begin{equation}\label{c3a}
\begin{split}
& \epsilon \phi_2^{(2, c)}=v_1+\phi_3^{(2, a)}, \\
& \epsilon^2 \phi_2^{(2, c)}=\epsilon v_1+\epsilon \phi_3^{(2, a)}, \\
& \epsilon^2 \phi_2^{(2, c)}=v_2+v_3+\epsilon \phi_3^{(2, a)}, \\
& \epsilon^3 \phi_2^{(2, c)}=\epsilon v_2+\epsilon v_3+\epsilon^2 \phi_3^{(2, a)}, \\
& \epsilon^3 \phi_2^{(2, c)}=\phi_2^{(2, c)}+\phi_3^{(2, c)}+\epsilon^2 \phi_3^{(2, a)}, \\
& \left(\epsilon^3-1\right) \phi_2^{(2, c)}=P_1(\epsilon) \phi_2^{(2, c)}=\phi_3^{(2, c)}+\epsilon^2 \phi_3^{(2, a)}.
\end{split}
\end{equation}
Then we can derive
\begin{equation}
\begin{split}
&P^2_1(\epsilon)\phi^{(2,c)}_1=\phi^{(2,c)}_3+\epsilon^2\phi^{(2,a)}_3,\\
&P^3_1(\epsilon)\phi^{(2,c)}_1=\phi^{(2,c)}_1+\epsilon^2\phi^{(2,a)}_1,\\
&[P_1^3(\epsilon)-1]\phi^{(2,c)}_1=P_2(\epsilon)\phi^{(2,c)}_1=\epsilon^2\phi^{(2,a)}_1.
\end{split}
\end{equation}
Similarly, we have
\begin{equation}
\begin{split}
&P_2(\epsilon)\phi^{(2,c)}_1=\epsilon^2\phi^{(2,a)}_1,\\
&P_2(\epsilon)\phi^{(2,a)}_1=\epsilon^2\phi^{(2,b)}_1,\\
&P_2(\epsilon)\phi^{(2,b)}_1=\epsilon^2\phi^{(2,c)}_1.
\end{split}
\end{equation}
It is straightforward to obtain the characteristic polynomial of $H_3$, which is $P_3(\epsilon)=P_2^3-(\epsilon^2)^3$. 

We then move on to the lattice with $n=4$ generations, where we have
\begin{equation}\label{c15}
\begin{split}
P_2(\epsilon)\phi^{(3,c)}_1&=\epsilon^2\phi^{(3,c)}_2,\\
P_2(\epsilon)\phi^{(3,c)}_3&=\epsilon^2\phi^{(3,c)}_1,\\
P_2(\epsilon)\phi^{(3,a)}_3&=\epsilon^2\phi^{(3,a)}_1,\\
P_2(\epsilon)\phi^{(3,c)}_2&=\epsilon^2\phi^{(3,c)}_3+P_1^2(\epsilon)\epsilon^2\phi^{(3,a)}_3.
\end{split}
\end{equation}
The derivation of the final equation in Eq. (\ref{c15}) is as follows:
\begin{equation}\label{c15a}
\begin{split}
& P_1(\epsilon) \phi_2^{(3, c)}=v_1+\epsilon^2 \phi_3^{(3, a)}, \\
& P_1(\epsilon) v_1=v_2+\epsilon^2 v_3, \\
& P_1(\epsilon) v_2=\phi_2^{(3, c)}, \\
& P_1(\epsilon) v_3=\phi_3^{(3, c)},\\
& P_1^2(\epsilon) \phi_2^{(3, c)}=P_1(\epsilon) v_1+P_1(\epsilon) \epsilon^2 \phi_3^{(3, a)}, \\
& P_1^2(\epsilon) \phi_2^{(3, c)}=v_2+\epsilon^2 v_3+P_1(\epsilon) \epsilon^2 \phi_3^{(3, a)}, \\
& P_1^3(\epsilon) \phi_2^{(3, c)}=P_1(\epsilon) v_2+\epsilon^2 P_1(\epsilon) v_3+P_1^2(\epsilon) \epsilon^2 \phi_3^{(3, a)}, \\
& P_1^3(\epsilon) \phi_2^{(3, c)}=\phi_2^{(3, c)}+\epsilon^2 \phi_3^{(3, c)}+P_1^2 \epsilon^2 \phi_3^{(3, a)}, \\
& \left(P_1^3(\epsilon)-1\right) \phi_2^{(3, c)}=P_2(\epsilon) \phi_2^{(3, c)}=\epsilon^2 \phi_3^{(3, c)}+P_1^2(\epsilon) \epsilon^2 \phi_3^{(3, a)}.
\end{split}
\end{equation}
Then we have
\begin{equation}
\begin{split}
&P_2^3(\epsilon)\phi^{(3,c)}_1=\epsilon^6\phi^{(3,c)}_1+P_1^2(\epsilon)\epsilon^6\phi^{(3,a)}_1,\\
&P_3(\epsilon)\phi^{(3,c)}_1=P_1^2(\epsilon)\epsilon^6\phi^{(3,a)}_1.
\end{split}
\end{equation}
Therefore, we can determine the characteristic polynomial of $H_4$ as $P_4(\epsilon)=P_3^3(\epsilon)-(P^2_1(\epsilon)\epsilon^6)^3$.

Generally, for the case of a lattice with $n$ generations, 
\begin{equation}\label{c15b}
\begin{split}
P_{n-2}(\epsilon)\phi^{(n-1,c)}_1&=\prod_{i=0}^{n-4} P_i^2(\epsilon)\left[P_{n-3}^3(\epsilon)-P_{n-2}(\epsilon)\right]^{\frac23}\phi^{(n-1,c)}_2,\\
P_{n-2}(\epsilon)\phi^{(n-1,c)}_3&=\prod_{i=0}^{n-4} P_i^2(\epsilon)\left[P_{n-3}^3(\epsilon)-P_{n-2}(\epsilon)\right]^{\frac23}\phi^{(n-1,c)}_1,\\
P_{n-2}(\epsilon)\phi^{(n-1,a)}_3&=\prod_{i=0}^{n-4} P_i^2(\epsilon)\left[P_{n-3}^3(\epsilon)-P_{n-2}(\epsilon)\right]^{\frac23}\phi^{(n-1,a)}_1,\\
P_{n-2}(\epsilon)\phi^{(n-1,c)}_2&=\prod_{i=0}^{n-4} P_i^2(\epsilon)\left[P_{n-3}^3(\epsilon)-P_{n-2}(\epsilon)\right]^{\frac23}\phi^{(n-1,c)}_3
+\prod_{i=0}^{n-3} P_{i}^2(\epsilon)\phi^{(n-1,a)}_3.
\end{split}
\end{equation}
Through direct derivations, we obtain the characteristic polynomials of $H_{n}$, as given by Eq. (9) in the main text.
